\documentclass[journal]{IEEEtran}

\ifCLASSINFOpdf
\else
\fi
\usepackage{amsmath}
\usepackage{amsmath}
\usepackage{amssymb} 
\usepackage{graphicx}
\usepackage{caption}
\usepackage{subcaption}
\usepackage{bm}
\usepackage{mathrsfs}
\usepackage{pdfrender}
\usepackage{tikz}
\usepackage{graphicx} 
\usepackage{changepage} 
\usepackage{algorithm}
\usepackage{algpseudocode}
\usepackage{algorithm}
\usepackage{algpseudocode}
\usepackage{booktabs} 
\usepackage{array}    
\usepackage{caption}  
\usepackage{cite}
\usepackage{bbm}
\usepackage[font=footnotesize]{caption}
\usepackage[font=footnotesize]{subcaption}
\usepackage{placeins}

\newcommand{\mathoutline}[1]{%
  \text{\pdfrender{StrokeColor=black, TextRenderingMode=1, LineWidth=0.08pt}{\ensuremath{#1}}}%
}

\usepackage{xcolor}
\usepackage{etoolbox}
\usepackage{tikz}

\newcommand*{\RedBibKeys}{}

\makeatletter
\newif\ifbib@incolor
\bib@incolorfalse

\usetikzlibrary{arrows.meta,positioning,shapes.geometric,calc}
\usetikzlibrary{arrows.meta,positioning,shapes.geometric,calc,fit,backgrounds}

\newcommand*{\bib@maybe@red}[1]{%
  \ifbib@incolor\endgroup\bib@incolorfalse\fi
  
  \def\found{0}%
  \def\do##1{%
    \ifstrequal{##1}{#1}{\def\found{1}}{}%
  }%
  \expandafter\docsvlist\expandafter{\RedBibKeys}%
  \ifnum\found=1\relax
    \begingroup\color{red}\bib@incolortrue
  \fi
}

\let\old@bibitem\@bibitem
\def\@bibitem#1{%
  \bib@maybe@red{#1}%
  \old@bibitem{#1}%
}

\let\old@lbibitem\@lbibitem
\def\@lbibitem[#1]#2{%
  \bib@maybe@red{#2}%
  \old@lbibitem[#1]{#2}%
}

\pretocmd{\endthebibliography}{%
  \ifbib@incolor\endgroup\bib@incolorfalse\fi
}{}{}

\makeatother

\begin{document}

\title{Hierarchical Finite-Element Analysis of Multiscale Electromagnetic Problems via Sparse Operator-Adapted Wavelet Decomposition\\}

\author{F.~\c{S}\i k,~\IEEEmembership{Graduate Student Member,~IEEE,}
        F.~L.~Teixeira,~\IEEEmembership{Fellow,~IEEE,}
        and~B.~Shanker,~\IEEEmembership{Fellow,~IEEE}
\thanks{Manuscript received 27 July 2025; revised 18 January 2026; accepted 12 February 2026.}
\thanks{The authors are with the Department of Electrical and Computer Engineering and Electroscience Laboratory, The Ohio State University, Columbus, OH 43210, USA (\textit{Corresponding author: Furkan \c{S}\i k})}}

\maketitle


\begin{abstract}
In this paper, we present a finite element method (FEM) framework enhanced by an operator-adapted wavelet decomposition algorithm designed for the efficient analysis of multiscale electromagnetic problems. Usual adaptive FEM approaches, while capable of achieving the desired accuracy without requiring a complete re-meshing of the computational domain, inherently couple different resolution levels. This coupling requires recomputation of coarser-level solutions whenever finer details are added to improve accuracy, resulting in substantial computational overhead. Our proposed method addresses this issue by \emph{decoupling} the different resolution levels. This feature enables independent computations at each scale that can be incorporated into the solutions to improve accuracy whenever needed, without requiring re-computation of coarser-level solutions. The main algorithm is hierarchical, constructing solutions from finest to coarser levels through a series of sparse matrix-vector multiplications. Due to its sparse nature,  the overall computational complexity of the algorithm is nearly linear. Moreover, Krylov subspace iterative solvers are employed to solve the final linear equations, with ILU preconditioners that enhance solver convergence and maintain overall computational efficiency. The proposed algorithm is applicable to both structured and unstructured meshes. Several two-dimensional numerical experiments demonstrate its high precision and nearly \(\mathcal{O}(N)\) computational complexity.
\end{abstract}


\section{Introduction}
\IEEEPARstart{T}{he} finite element method (FEM) is instrumental in obtaining accurate and reliable numerical solutions to complex electromagnetic problems. Ongoing theoretical advancements have further strengthened its capabilities, leading to widespread adoption~\cite{jin2015finite,peterson1998computational}. Despite its precision and efficiency in solving boundary value problems (BVPs) derived from Maxwell's equations, the application of FEM to problems involving sharp corners, high field gradients, singularities, geometric discontinuities, and multiscale features in general still has some challenges. When (uniform) meshes are sufficiently fine to capture small-scale features, typically confined to small regions of the domain, the remainder of the domain becomes over-resolved; conversely, insufficient mesh refinement sacrifices accuracy in regions of high field gradients~\cite{quraishi2009adaptive, filippi2015performance, filippi2020electromagnetic}. \emph{Adaptive} mesh refinement strategies address these issues by refining the mesh only when needed, thus avoiding the remeshing of the entire domain \cite{vanti1997optimal, filippi2015performance, filippi2020electromagnetic}. However, the resulting FEM matrices couple different resolution levels, forcing a new re-computation of the entire solution (including previously computed coarser levels). In addition, they typically show a degradation in their condition number, with a consequent slowdown of iterative solutions. To remove these drawbacks, it is highly desirable to develop multiscale FEM algorithms with no coupling between resolution levels~\cite{vanti1997optimal, filippi2015performance, filippi2020electromagnetic}.

Wavelets provide a natural multiscale representation of functions that supports hierarchical approximations to exact solutions. Because wavelets typically possess compact support, they are well suited for describing localized phenomena such as sharp field gradients and singularities \cite{mallat1989theory, daubechies1990wavelet, williams1994introduction}. The earliest attempts to integrate wavelets into the FEM primarily consisted of employing (first-generation) bi-orthogonal wavelets as basis functions. However, similar to conventional adaptive FEM, these first-generation wavelets produce scale-coupled stiffness matrices, thus preventing the development of efficient scale-decoupled systems \cite{harbrecht2002biorthogonal, he2007adaptive, wang2012second, williams1994introduction,7588624}. These obstacles spurred the development of second-generation wavelets that achieved scale-decoupled stiffness matrices. However, the second-generation wavelet-based FEM approaches explored in the literature face challenges when generalizing to different mesh types, geometries, and boundary conditions. Moreover, these algorithms typically incur higher overall computational costs, especially in problems involving unstructured meshes~\cite{he2007adaptive, wang2012second, amaratunga2006multiresolution, filippi2020electromagnetic, sudarshan2006combined, quraishi2011second, 7588624, hughes1998vms, hou1997msfem, ji2015multiscale_em_ctp, giraldo2023variationalmultiscalemethodderived}. 

More recently, \emph{operator-adapted} wavelet decomposition algorithms based on a hierarchical linear-algebraic construction that produced scale-decoupled FEM systems have been developed~\cite{budninskiy2019operator, chen2019material, owhadi2019operator, sudarshan2005operator}. Similarly to other wavelet-based algorithms, operator-adapted wavelet algorithms expand the solution into a coarse level and a hierarchy of detail levels that can locally capture fine details of the solution. A unique feature of operator-adapted wavelets is \emph{full decoupling} among the coarse level and all detail levels. Because of this decoupling, once the coarser-level solution is obtained, there is no need to re-evaluate it when adding finer-level details. Further, it is also unnecessary to recompute finer level details when adding deeper levels of detail. After computing the coarser-level FEM solution, one can easily add subsequent detail levels by solving small, independent linear equations to improve solution accuracy. Moreover, each detail function inherently indicates where higher resolution is needed based on its magnitude, eliminating the need for a separate error estimation step. 

Our contributions in this paper are three fold; (i) we  extend operator-adapted wavelets to unstructured meshes, (ii) we present a near-linear scaling approach to both constructing and solving the systems of equations that arise for unstructured meshes, and (iii) we validate the applicability and the nearly linear computational complexity of the proposed method. Specifically, to date, operator-adapted basis functions have been developed for structured meshes, where obtaining a (near) linear scaling algorithm is fairly straightforward. In this paper, we develop a new operator-adapted wavelet FEM algorithm for irregular triangular elements at all scale levels and vector edge basis functions (Whitney one forms)~\cite{bossavit1989,teixeira2014} suitable for EM applications. As a result, developing a linear scaling algorithm is significantly more challenging. In fact, direct application of the methods developed in {\cite{budninskiy2019operator,chen2019material}} leads to the cost complexity of $\mathcal{O}(N^3)$. To address this issue, we develop new strategies to construct sparse representations at all levels and judiciously exploit sparse linear algebra tools to obtain near \( \mathcal{O}(N) \) complexity. We are able to eliminate the use of dense intermediate matrices by leveraging a hierarchical coarsening procedure, sparse matrix–vector multiplications, and other sparsity-based operations.

The nearly \(\mathcal{O}(N)\) computational complexity is verified through simulations of several two-dimensional multiscale electromagnetic problems, including waveguide discontinuities (L-shaped and U-shaped geometries with sharp corners and high field gradients~\cite{garcia2007two}), and a leaky waveguide comprised of two parallel microporous Si (MPSi) slabs. The results are compared against conventional (finest-level) FEM solutions and, where available, analytical or semi-analytical benchmark solutions.

The remainder of the paper is organized as follows. Section \ref{sec:theory} presents the problem statement, the general finite element discretization, and a brief overview of traditional wavelet-Galerkin methods. Section \ref{sec:opFEM} reviews canonical multiresolution analysis and how to obtain precomputed, operator-agnostic refinement matrices. It also presents a brief discussion of the explicit matrix-based version of the operator-adapted wavelet decomposition algorithm. Section \ref{sec:opEfficient} discusses our novel algorithm, from precomputation to computing final solutions at each scale level, using sparse linear algebra-based strategies. Several numerical experiments demonstrating the accuracy and computational efficiency of the proposed algorithm are provided in Section~\ref{sec:res}. Finally, Section \ref{sec:conclusion} summarizes the contributions of this paper and suggests possible venues for future work.

\section{Theoretical Background\label{sec:theory}}
\subsection{Problem Statement}
Consider boundary value problems (BVPs) represented by equations of the form
\begin{equation}
\mathcal{L} \mathbf{u} = \mathbf{g}
\end{equation}
where $\mathcal{L}$ represents a tensor differential operator that incorporates both the differential equation and the boundary conditions. The vector $\mathbf{u}$ is the unknown vector function to be determined, and $\mathbf{g}$ is a given vector function. Let $\Omega$ denote a domain whose boundary $\partial \Omega$  is $\Gamma = \bigcup_i \Gamma_i$. The Helmholtz vector equation will be considered in the numerical experiments of this paper, with \(\mathbf{u}(\mathbf{r})\) representing the electric field or the magnetic field intensity:
\begin{equation}
\left( \nabla \times \left[ \frac{1}{\alpha} \nabla \times \right] - k_0^2 \theta\right) \mathbf{u}(\mathbf{r}) = \mathbf{g}(\mathbf{r}) \quad \text{in} \quad \Omega
\end{equation}
with boundary conditions
\begin{equation}
\mathcal{B}_i \{ \mathbf{u}(\mathbf{r}) \} = \mathbf{b}_i(\mathbf{r}) \quad \text{on } \Gamma_i, 
\end{equation} represented by a generic differential operator
$\mathcal{B}_i$  and boundary data $\mathbf{b}_i(\mathbf{r})$ specified on $\Gamma_i$. Dirichlet, Neumann, or Robin boundary conditions can be applied as appropriate.
The parameters \(\alpha\) and \(\theta\) indicate the material properties: if \(\mathbf{u}(\mathbf{r})\) is the electric field intensity, then (\(\alpha\),\(\theta\)) refers to (\(\mu_r\),\(\epsilon_r\)); conversely, if \(\mathbf{u}(\mathbf{r})\) is the magnetic field intensity, then
(\(\alpha\),\(\theta\)) refers to (\(\epsilon_r\),\(\mu_r\)). Moreover, \(\mathbf{g}(\mathbf{r})\) corresponds to the source term and \(k_0\) is the wavenumber in free-space. 

\subsection{Finite Element Discretization}

In the numerical experiments presented in this paper, a two-dimensional (2D) domain \( \Omega \) is subdivided into triangular elements to discretize the problem. Each element covers a subdomain \( \Omega^e \), for \( e = 1, 2, \ldots, n^e \), where \( n^e \) denotes the total number of elements. The unknown function within each element, \( \mathbf{u}^e(\mathbf{r}) \), is expressed as a weighted sum of basis functions
\begin{equation}
\mathbf{u}^e(\mathbf{r}) = \sum_{j=1}^{s^e} u_j^e \mathbf{W}_j^e(\mathbf{r}),
\end{equation}
where \(u_j^e\) represents the unknown degrees of freedom (DoF) associated with the \(j\)th basis function, \( s^e \) is the number of DoFs in each element (\( s^e = 3 \) for triangular elements), and \(\mathbf{W}_j^e(\mathbf{r})\) are the chosen basis functions. As customary, Whitney one-forms (edge elements) are used as the basis functions in a Sobolev space, \(\mathbf{W}_j^e(\mathbf{r}) \in H(\mathrm{curl}, \Omega^e)\).

Before detailing the operator-adapted wavelet-Galerkin method developed in this work, the next subsection introduces the classical use of wavelet basis functions and the traditional wavelet-Galerkin approach, providing the necessary definitions.

\subsection{Multiresolution Analysis and Wavelet-Galerkin Method}
The traditional wavelet-Galerkin method uses a compactly supported, finite-energy orthogonal basis to solve (1). This method  approximates the solution by constructing a finite-dimensional subspace \( V^q \) and, within a multiresolution framework, forms a nested sequence of approximation spaces \( \{ V^j \}_{j=1}^{q-1} \subset H \). The hierarchy of approximation spaces at different resolution levels can be expressed as
\begin{equation}
\{ 0 \}\subset V^1\subset\cdots\subset V^{{j-1}}\subset V^{j}\subset V^{{j+1}}\subset\cdots \subset V^q.
\end{equation}
For each approximation space (solution space) \( \{ V^j \}_{j=1}^{q-1} \), there exists a complementary wavelet space \( W^j \) that contains the ``details" needed to transition from \( V^j \) to the finer space \( V^{j+1} \). These wavelet spaces (detail spaces) satisfy:
\begin{equation}
{V}^{j+1} = {V}^j \oplus{W}^j \quad \text{for } j = 1,2, \dots, q-1
\end{equation}
where $\oplus$ denotes the direct sum operation in the $L^2$ space. Using this hierarchical approach, the projection of the function $\mathbf{u}$  onto the solution space \( V^{j+1} \), that is \( \mathbf{u}^{j+1} \), is given as
\begin{equation}
\mathbf{u}^{j+1} = \mathbf{u}^1 + \sum_{i=1}^{j} \mathbf{d}^i,
\end{equation}
where $\mathbf{u}^1$ is the projection of $\mathbf{u}$  onto the coarsest space \( V^1 \), and $\mathbf{d}^i$ are the detail functions corresponding to the wavelet spaces \( W^i \) up to level \( j \).
By recursively decomposing the solution spaces into coarser levels and their corresponding detail spaces until reaching the coarsest level, the solution space \( V^q \) admits the following multiresolution decomposition:
\begin{equation}
H \approx V^q = V^1 \oplus W^1 \oplus W^2 \oplus \cdots \oplus W^{q-1}.
\end{equation}
Each approximation space function $\mathbf{u}^{j}$  can be represented as \( \mathbf{u}^{j} = \sum_{i=1}^{n_j} v_i^j  \boldsymbol{\phi}_i^j\), where the set \( \{\boldsymbol{\phi}_i^j\}_{i=1}^{n_j} \) forms a basis of \( V^j \). Similarly, \( \{\boldsymbol{\psi}_\ell^j\}_{\ell=1}^{N_j}\) forms a basis of the wavelet space \( W^j \), with \(N_j=n_{j+1}-n_j\). 
Consequently, the solution \( \mathbf{u}^q \) can be expressed as follows:

\begin{equation}
\mathbf{u}^q = \sum_{i=1}^{n_1} v_i^1 \boldsymbol{\phi}_i^1 + \sum_{j=1}^{q-1} \sum_{\ell=1}^{N_j} w_\ell^j \boldsymbol{\psi}_\ell^j.
\end{equation}
To find the solution represented by (9) using a traditional wavelet-Galerkin FEM method, the following linear system can be solved:
\begin{equation}
\mathbf{L} \, \textbf{u} = \textbf{g},
\end{equation}
where the global FEM matrix \( \mathbf{L} \) has size
\( n_q \times n_q \), with   \( n_q = n_1 + \sum_{j=1}^{q-1} N_j\)  being the the total number of degrees of freedom for \(V^q\). The matrix
\( \mathbf{L} \)  is composed of sub-matrices of the form:
\begin{equation}
{\mathbf{L}} =
\left[
\begingroup
\renewcommand{\arraystretch}{1.5} 
\begin{array}{ccccc}
A^1 & M_{\phi \psi}^{(1,1)} & M_{\phi \psi}^{(1,2)} & \cdots & M_{\phi \psi}^{(1,q-1)} \\
M_{\psi \phi}^{(1,1)} & B^1 & M_{\psi \psi}^{(1,2)} & \cdots & M_{\psi \psi}^{(1,q-1)} \\
M_{\psi \phi}^{(2,1)} & M_{\psi \psi}^{(2,1)} & B^2 & \cdots & M_{\psi \psi}^{(2,q-1)} \\
\vdots & \vdots & \vdots & \ddots & \vdots \\
M_{\psi \phi}^{(q-1,1)} & M_{\psi \psi}^{(q-1,1)} & M_{\psi \psi}^{(q-1,2)} & \cdots & B^{q-1}
\end{array}
\endgroup
\right]
\end{equation}

The sub-matrix entries of \( {\mathbf{L}} \) are defined as follows:

\begin{itemize}
 \setlength\itemsep{0.3em} 
\item \( A^1 =  A_{i\ell}^1 := L(\boldsymbol{\phi}_i^1, \boldsymbol{\phi}_\ell^1) \) is the \( n_1 \times n_1 \) stiffness matrix of the coarsest level, \(V^1\) 

\item \( B^j =  B_{i\ell}^j := L(\boldsymbol{\psi}_i^j, \boldsymbol{\psi}_\ell^j) \) is the \( N_j \times N_j \) = \( (n_{j+1}-n_j) \times (n_{j+1}-n_j) \) stiffness matrix of the detail level, \(W^j\)

\item \( M_{\boldsymbol{\phi} \boldsymbol{\psi}, \, i\ell}^{(1,t)} = L(\boldsymbol{\phi}_i^1, \boldsymbol{\psi}_\ell^t) = M_{\boldsymbol{\psi} \boldsymbol{\phi}, \, {\ell}i}^{(t,1)} \) represents the coupling terms between different detail spaces and the first approximation (solution) space 

\item \( M_{\boldsymbol{\psi} \boldsymbol{\psi}, \, i\ell}^{(s,t)} = L(\boldsymbol{\psi}_i^s, \boldsymbol{\psi}_\ell^t) \), for \( s \neq t \) represents the interactions across wavelet bases at different levels s and t.
\end{itemize}
Furthermore, the right-hand side vector \( {\mathbf{g}} \) and the coefficient vector \( {\mathbf{u}} \) are defined in a similar hierarchical structure:

\[
\mathbf{g} = \begin{bmatrix}
    \mathbf{g}^1 \\
    \mathbf{b}^1 \\
    \vdots \\
    \mathbf{b}^{q-1}
\end{bmatrix}
\quad \text{and} \quad
\mathbf{u} = \begin{bmatrix}
    \mathbf{v}^1 \\
    \mathbf{w}^1 \\
    \vdots \\
    \mathbf{w}^{q-1}
\end{bmatrix}
\]
where $\mathbf{g_i^1}=G(\boldsymbol{\phi}_i^{1})
=\langle \mathbf{g},\boldsymbol{\phi}_i^{1}\rangle
=\int_{\Omega}\mathbf{g}\cdot\boldsymbol{\phi}_i^{1}\,\mathrm{d}\Omega$
and detail-level right-hand-sides $b_\ell^{j}=G(\boldsymbol{\psi}_\ell^{j})
=\langle \mathbf{g},\boldsymbol{\psi}_\ell^{j}\rangle
=\int_{\Omega}\mathbf{g}\cdot\boldsymbol{\psi}_\ell^{j}\,\mathrm{d}\Omega$,
for $j=1,\dots,q-1$.

The off-diagonal sub-matrices in the stiffness matrix \({\mathbf{L}} \) represent the coupling between different resolution levels. Non-zero coupling terms indicate that the coarser-level solution is influenced by the addition of detail levels. In designing adaptive multiscale solvers, it is highly desirable for the details to not influence the coarser-level solution. However, in traditional wavelet-Galerkin method this is not the case because many off-diagonal terms are non-zero. In addition, the condition number of the matrices deteriorates as \( q \) increases. These issues cause traditional wavelet-Galerkin approach not to show a meaningful advantage over conventional FEM refinement methods \cite{budninskiy2019operator, vanti1997optimal, filippi2020electromagnetic, filippi2015performance, amaratunga2006multiresolution, sudarshan2005operator, sudarshan2006combined, chen2019material, owhadi2019operator}. 
To eliminate undesirable coupling across scales, customized wavelets have been recently developed~\cite{owhadi2019operator, budninskiy2019operator, chen2019material}, as discussed in the next section.

\section{Operator-Adapted Wavelet-Based FEM \label{sec:opFEM}}

This section introduces the operator-adapted wavelet decomposition algorithm, describing its key features. In particular, the need for a sparsity-driven, efficient reformulation is emphasized to motivate the novel operator-adapted wavelet decomposition algorithm described in Section IV. Additionally, precomputed operator-agnostic refinement matrices, derived from usual FEM basis functions, are described along with a novel efficient construction that leverages locality.

\subsection{Preliminaries}

\subsubsection{Notation}

 In this paper, precomputed operator-agnostic matrices, vectors, and variables, which are obtained using usual FEM procedure or basis functions, are denoted by bold letters with a tilde on top e.g., \(\tilde{\mathbf{C}}^j\). In contrast, operator-adapted matrices, vectors, and variables, generated by the proposed operator-adapted wavelet decomposition algorithm, are represented in outlined fonts e.g., \(\mathbb{C}^j\).

\subsubsection{Definition of Operator-Adapted Wavelets and Functional Spaces}

The operator-adapted wavelet space \(\mathbb{W}^j \) at any resolution level \( j \) is operator-orthogonal (\( \mathcal{L} \)-orthogonal) to the solution space at its own resolution level and to the wavelet spaces at all other levels. Thus, (6), which is expressed for traditional multiresolution analysis (MRA), can be reformulated within the framework of operator-orthogonality as follows:
\begin{equation}
{\mathbb{V}}^{j+1} = {\mathbb{V}}^j \oplus_{\mathcal{L}}{\mathbb{W}}^j \quad \text{for } j = 1,2, \dots, q-1,
\end{equation}
where \( \oplus_{\mathcal{L}} \) denotes the direct sum of \( \mathcal{L} \)-orthogonal subspaces. In this setting, $\mathbb{W}^{j}$ is spanned by the operator-adapted wavelets $\{\mathoutline{\psi}_\ell^j\}_{\ell=1}^{N_j}$, whereas the operator-adapted basis functions $\{\mathoutline{\phi}_i^j\}_{i=1}^{n_j}$ span the solution space $\mathbb{V}^{j}$. The hierarchy for the finest-level solution space \( \mathbb{V}^q \) can be constructed using the operator-adapted multiresolution approach as follows:
\begin{equation}
H \approx {\mathbb{V}}^{q} = {\mathbb{V}}^{1} \oplus_{\mathcal{L}} {\mathbb{W}}^{1} \oplus_{\mathcal{L}} \dots \oplus_{\mathcal{L}} {\mathbb{W}}^{q-2} \oplus_{\mathcal{L}} {\mathbb{W}}^{q-1}.
\end{equation}
Using this approach, a scale-decoupled system is obtained; the inter-level stiffness matrices have vanished, i.e., \( M^{(s,t)} = 0 \) for all \( s, t \). Thus, the entire system can be analyzed by solving \( q \) independent linear equations, and (11) becomes:

\begin{equation}
{\mathbb{L}} =
\left[
\begingroup
\renewcommand{\arraystretch}{1.5} 
\begin{array}{ccccc}
{\mathbb{A}}^1 & 0 & \cdots & 0 \\
0 & {\mathbb{B}}^1  & \cdots & 0 \\
\vdots & \vdots & \ddots & \vdots \\
0 & 0 & \cdots & {\mathbb{B}}^{q-1}
\end{array}
\endgroup
\right].
\end{equation}

To find the unknown coefficients, \({\mathbbm{u}} =\begin{bmatrix} \mathbbm{v}^{1} 
\mathbbm{w}^1 \!\cdots \!\mathbbm{w}^{q-1} \end{bmatrix}^T\), 
using the operator-adapted wavelet decomposition algorithm, the equation \({\mathbb{L}} \, {\mathbbm{u}} = {\mathbbm{g}}\) will be solved using the stiffness matrix provided in (14) and the right-hand side vector, \({\mathbbm{g}} =
\begin{bmatrix} \mathbbm{g}^1 \mathbbm{b}^1 \!\cdots \!\mathbbm{b}^{q-1} \end{bmatrix}^T\).

\subsubsection{Canonical Multiresolution Analysis}

The proposed operator-adapted wavelet decomposition method is constructed on a nested sequence of refinable basis functions \(\{\tilde{\boldsymbol{\phi}}_i^j\}_{i=1}^{n_j}\), which span a hierarchy of multiscale approximation spaces \(V^j = \text{span}\{\tilde{\boldsymbol{\phi}}_i^j\}_{i=1}^{n_j}\). Each \(V^j\) corresponds to a mesh \( \mathcal{M}^j \) at resolution level \(j\), with meshes becoming progressively finer over the domain $\Omega$.
For \(1 \leq j < q\), each basis function \(\tilde{\boldsymbol{\phi}}_i^j\) in \(V^j\) can be represented as a weighted sum of basis functions \(\tilde{\boldsymbol{\phi}}_\ell^{j+1}\) in \(V^{j+1}\)
\begin{equation}
\tilde{\boldsymbol{\phi}}_i^j = \sum_{\ell=1}^{n_{j+1}} \tilde{\mathbf{C}}_{i\ell}^j \, \tilde{\boldsymbol{\phi}}_\ell^{j+1}
\end{equation}
where \(\tilde{\mathbf{C}}^j\) is a sparse, operator-agnostic refinement matrix between successive levels, constructed using user-specified operator-agnostic basis functions.

The initial matrices for setting up the operator-adapted wavelet decomposition algorithm include the operator-agnostic refinement matrices and their null spaces across all resolution levels. The null-space matrix (refinement kernel matrix) can be computed as follows: 

\begin{equation}
\tilde{\mathbf{C}}^j \tilde{\mathbf{W}}^{j, T} = \mathbf{0}_{n_j \times N_j}
\end{equation}

In the following section, we will \emph{explain} how \(\tilde{\mathbf{W}}\) can be computed with a nearly linear computational complexity. Furthermore, the condition number of this sparse null-space matrix \(\tilde{\mathbf{W}}\) is very close to 1. In addition to the operator-agnostic refinement matrices and their null spaces across all resolution levels, edge basis functions $\{\tilde{\boldsymbol{\phi}}^{\,q}_i\}_{i=1}^{n_q}$, usual FEM basis function matrix $\tilde{\boldsymbol{\Phi}}^{q}
=
\big[\tilde{\boldsymbol{\phi}}^{\,q}_{1}\ \tilde{\boldsymbol{\phi}}^{\,q}_{2}\ \cdots\ \tilde{\boldsymbol{\phi}}^{\,q}_{n_q}\big]^{T}$, conventional FEM stiffness matrices, and the right-hand side vectors generated at the finest resolution level are directly used in the operator-adapted wavelet decomposition algorithm:

\begin{equation}
\tilde{\mathbf{A}}^q = {\mathbb{A}}^q, \quad \tilde{\bm{\Phi}}^q = \mathoutline{\Phi}^q, \quad \tilde{\mathbf{g}}^q = {{\mathbbm{g}}}^q.
\end{equation}

\subsubsection{Matrix-Based Algorithm}
An overview of the key concepts of the matrix-based algorithm that serves as the foundation of our new reformulation is provided here. More details on the matrix-based algorithm are available in references~\cite{owhadi2019operator,chen2019material,budninskiy2019operator}. As discussed previously, operator-adapted wavelets are constructed through a sequence of linear equations, progressing hierarchically from finer to coarser levels. At each level \( j = 1, \dots, q \), we construct \( n_j \) operator-adapted basis functions, denoted \( \{ \mathoutline{\phi}_i^j \}_{i=1}^{n_j}\), which span the approximation space \(\mathbb{V}^j = \text{span}\{\mathoutline{\phi}_i^j\}_{i=1}^{n_j}\). The operator-adapted coarser-level basis functions \(\mathoutline{\phi}_i^{j}\) can be expressed in terms of the finer-level functions \(\mathoutline{\phi}_\ell^{j+1}\):
\begin{equation}
\mathoutline{\phi}_i^j = \sum_{\ell=1}^{n_{j+1}} {\mathbb{C}}_{i\ell}^j \, \mathoutline{\phi}_\ell^{j+1}.
\end{equation}
where \({\mathbb{C}}^j\) denotes the \(n_j \times n_{j+1}\) operator-adapted refinement matrix. Also, operator-adapted wavelets \( \{ \mathoutline{\psi}_i^j \}_{i=1}^{N_j}\) satisfy:
\begin{equation}
\mathoutline{\psi}_i^j = \sum_{\ell=1}^{n_{j+1}} \tilde{\mathbf{W}}_{i\ell}^j \, \mathoutline{\phi}_\ell^{j+1}.
\end{equation}

At each resolution level, the intermediate matrices required for the matrix-based algorithm are constructed from operator-agnostic matrices and vectors, combined with the operator-adapted basis functions and wavelets described in (18) and (19). Taking into account the scale-decoupled stiffness matrix in (14), the level-\(\zeta\) coefficients, i.e. \(\mathbbm{v}^{1}\) and \(\mathbbm{w}^j\) for \(j = 1, 2, \ldots, \zeta-1\), are determined by solving the following linear equations:
\begin{align}
\text{For coarsest-level:} \quad & {\mathbb{A}}^1 {\mathbbm{v}}^1 = {\mathbbm{g}}^1, \notag \\
\text{For detail-levels:} \quad & {\mathbb{B}}^j {\mathbbm{w}}^j = {\mathbbm{b}}^j, \quad \text{for } j = 1, \ldots, \zeta-1.
\end{align}

It is important to note that in this matrix-based approach the intermediate matrices at each level are obtained using computationally expensive operations, including matrix-matrix multiplications and matrix inversions. Using these operations and matrices, the linear system described in (20) is obtained. Furthermore, by using intermediate matrices whose details can be found in references~\cite{owhadi2019operator,chen2019material,budninskiy2019operator} the solution of (10) at scale level \(\zeta\) can be obtained as:

\begin{equation}
\mathbbm{u}^\zeta  = \sum_{i=1}^{n_1} \mathbbm{v}_i^1 \mathoutline{\phi}_i^1 
+ \sum_{j=1}^{\zeta-1} \sum_{\ell=1}^{N_j} \mathbbm{w}_{\ell}^j \mathoutline{\psi}_{\ell}^j.
\end{equation}

Due to the independence of the linear equations in (20), the solution at any scale level can be determined by solving only the first \(\zeta\) independent linear equations. The stiffness matrix at the coarsest level \({\mathbb{A}}^1\) and the right side \(\mathbbm{{g}}^1\), along with the stiffness matrices at the detail level \({\mathbb{B}}^j\) and the right side \({\mathbbm{b}}^j\), are hierarchically calculated from finer to coarser levels. Using these calculations, the solution given in (21) is obtained.

\subsection{Computation of Operator-Agnostic Refinement Matrices}
In this subsection, operator-agnostic refinement matrices \(\{\tilde{\mathbf{C}}^j\}_{j=1}^{q-1}\), representing the weighted sum coefficients between operator-agnostic basis functions at resolution levels \(j\) and \(j+1\), will be efficiently calculated for edge basis functions on 2D triangular elements. Indeed, the operator-adapted wavelet decomposition method utilizes refinable basis functions at different scale levels independent of the mesh; thus, any mesh hierarchy can be employed. The mesh hierarchy used in our numerical experiments at various scale levels are illustrated in Fig.~1. Although the operator-adapted wavelet decomposition algorithm is a coarsening procedure developed from finer to coarser scales, the mesh hierarchy can be constructed using a subdivision-based approach for simplicity. 

To efficiently obtain the operator-agnostic refinement matrix $\tilde{\mathbf{C}}^j$, we compute the subdivision matrix, $\tilde{\mathbf{R}}^j$, between successive scale levels. Using the compact support of edge basis functions, the subdivision matrix can be obtained using local inner product vectors and local Gram matrices. As an example, for each coarser edge c, the local inner product vector ${\tilde{\mathbf{I}}}_{local}$ and the local Gram matrix ${\tilde{\mathbf{G}}}_{local}$ for the coarser-finer level basis function pairs ($\tilde{\boldsymbol{\phi}}_{c}^j$, $\tilde{\boldsymbol{\phi}}_{f}^{j+1}$) can be computed as follows:

\begin{equation}
\renewcommand{\arraystretch}{2} 
\setlength{\arraycolsep}{3pt} 
{\tilde{\mathbf{I}}_{\textit{local}}} = \begin{bmatrix}
\langle \tilde{\boldsymbol{\phi}}_{c}^j, \tilde{\boldsymbol{\phi}}_{1}^{j+1} \rangle \\
\langle \tilde{\boldsymbol{\phi}}_{c}^j, \tilde{\boldsymbol{\phi}}_{2}^{j+1} \rangle \\
\vdots \\
\langle\tilde{\boldsymbol{\phi}}_{c}^j, \tilde{\boldsymbol{\phi}}_{\gamma}^{j+1} \rangle
\end{bmatrix}
\end{equation}

and

\begin{equation}
{\tilde{\mathbf{G}}_{\textit{local}}} =
\left[
\begingroup
\renewcommand{\arraystretch}{2} 
\setlength{\arraycolsep}{3pt} 
\begin{array}{cccc}
\langle \tilde{\boldsymbol{\phi}}_1^{j+1}, \tilde{\boldsymbol{\phi}}_1^{j+1} \rangle & \langle \tilde{\boldsymbol{\phi}}_1^{j+1}, \tilde{\boldsymbol{\phi}}_2^{j+1} \rangle & \cdots & \langle \tilde{\boldsymbol{\phi}}_1^{j+1}, \tilde{\boldsymbol{\phi}}_{{\gamma}}^{j+1} \rangle \\
\langle \tilde{\boldsymbol{\phi}}_2^{j+1}, \tilde{\boldsymbol{\phi}}_1^{j+1} \rangle & \langle \tilde{\boldsymbol{\phi}}_2^{j+1}, \tilde{\boldsymbol{\phi}}_2^{j+1} \rangle & \cdots & \langle \tilde{\boldsymbol{\phi}}_2^{j+1}, \tilde{\boldsymbol{\phi}}_{\gamma}^{j+1} \rangle \\
\vdots & \vdots & \ddots & \vdots \\
\langle \tilde{\boldsymbol{\phi}}_{\gamma}^{j+1}, \tilde{\boldsymbol{\phi}}_1^{j+1} \rangle & \langle \tilde{\boldsymbol{\phi}}_{\gamma}^{j+1}, \tilde{\boldsymbol{\phi}}_2^{j+1} \rangle & \cdots & \langle \tilde{\boldsymbol{\phi}}_{\gamma}^{j+1}, \tilde{\boldsymbol{\phi}}_{\gamma}^{j+1} \rangle
\end{array}
\endgroup
\right]
\end{equation}
$ \text{for } j = 1, 2, 3, \ldots, q-1$. The \(\gamma \times 1\) local subdivision vectors $\tilde{\mathbf{{R}}}_{\text{\textit{local}}}$ for each coarser edge \(c\) are obtained by solving the following linear system:

\begin{equation}
\tilde{\mathbf{{I}}}_{\text{\textit{local}}} = \tilde{\mathbf{{G}}}_{\text{\textit{local}}} \, \tilde{\mathbf{{R}}}_{\text{\textit{local}}},
\end{equation}
where \(\tilde{\mathbf{{I}}}_{\text{\textit{local}}}\) is a \(\gamma \times 1\) local inner product vector and \(\tilde{\mathbf{{G}}}_{\text{\textit{local}}}\) is a \(\gamma \times \gamma\) local Gram matrix. The numerical integrations required to compute the inner products given in (22) and (23) are evaluated via Gaussian quadrature \cite{akin2005finite}.

This process is performed for each coarser mesh edge. The dimension parameter \(\gamma\) of the local matrices and vectors corresponds to the total number of finer-level edges that interact with the selected coarser edge within its associated triangles; \(\gamma\) = 16 for interior edges, as the coarser edge is associated with two triangles, and 
\(\gamma\) = 9 for boundary edges, as the coarser edge is associated with only one triangle. Once all local vectors are computed for each coarser edge, the global subdivision matrix $\tilde{\mathbf{R}}^j$ is constructed via local-to-global mapping. $\tilde{\mathbf{R}}^j$ is highly sparse, with only a few non-zero entries in each row and column, as each coarser edge interacts only with the finer edges in its neighboring triangles. This sparsity and localized nature of the method ensures that the overall computational complexity remains close to linear, \(\mathcal{O}(N)\), where \(N\) is the number of edges. Experimental results verifying these findings will be presented in the subsequent sections. The operator-agnostic refinement matrix $\tilde{\mathbf{C}}^j$ is the transpose of the obtained global subdivision matrix $\tilde{\mathbf{R}}^j$:

\begin{equation}
\tilde{\mathbf{C}}^j = \tilde{\mathbf{R}}^{j, T}
\end{equation}

\begin{figure}[htbp]
    \centering
    \hfill
    \begin{subfigure}[b]{0.4\textwidth}
        \centering
        \includegraphics[width=0.6\textwidth]{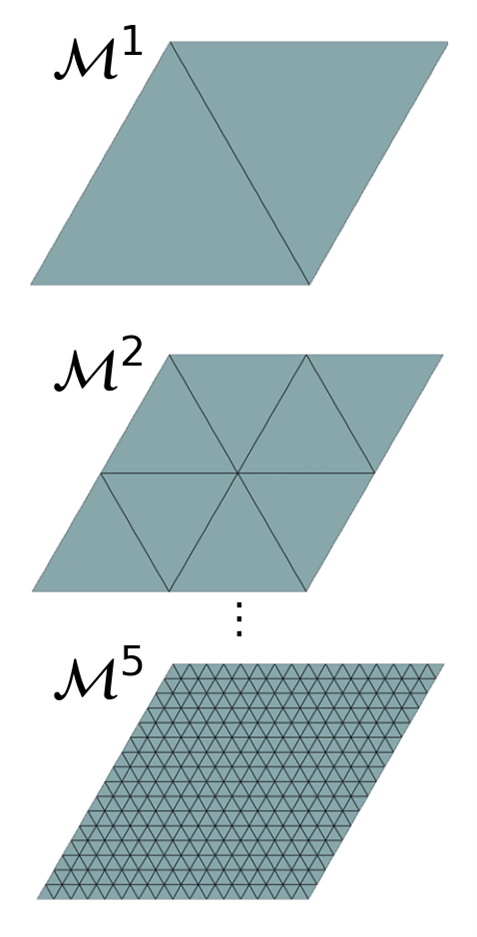}
    \end{subfigure}

    \caption{Mesh hierarchy for different resolution levels. $\mathcal{M}^1$: coarsest-level mesh, $\mathcal{M}^2$: second-level mesh after one subdivision, and $\mathcal{M}^5$: fifth-level mesh after four subdivisions.}
\end{figure}

\subsection{Sparse Linear Algebra Framework}

Although the operator-adapted wavelet decomposition approach yields a scale-decoupled system, maintaining near-linear computational complexity is crucial for large problems. However, the matrix-based approach described above relies on matrix–matrix multiplications, matrix inversions, and multiple linear solutions. Problematically, intermediate matrices typically exhibit a dense structure. As a result, almost all algorithmic steps involving these dense matrices incur cubic computational complexity. To overcome these challenges and enable solution of large problems involving unstructured grids and vector basis functions, we have reformulated the matrix-based algorithm leveraging sparse linear algebra tools and the hierarchical structure of the algorithm. This reformulation incorporates modifications inspired by the underlying physics and mathematical properties resulting in an algorithm with nearly \(\mathcal{O}(N)\) complexity. Building on this reformulation, our approach constructs the hierarchical structure solely from precomputed \emph{sparse} operator-agnostic matrices and vectors, thereby avoiding the computational burden associated with dense matrices.

\section{Efficient Construction of Operator-Adapted Wavelet Decompositions \label{sec:opEfficient}}

\subsection{Sparsity in Precomputed Operator-Agnostic Matrices}

In this subsection, we highlight the crucial role of sparsity in operator-agnostic matrices, which forms the foundation of our novel algorithmic reformulation for efficiently constructing the operator-adapted wavelet decomposition with nearly \( \mathcal{O}(N) \) complexity. The operator-agnostic refinement matrices \(\{\tilde{\mathbf{C}}^j\}_{j=1}^{q-1}\), their null space matrices \(\{\tilde{\mathbf{W}}^j\}_{j=1}^{q-1}\), the finest-level usual FEM stiffness matrix $\tilde{\mathbf{A}}^q$ and the finest-level usual FEM basis function matrix $\tilde{\bm{\Phi}}^q$ are all inherently sparse. This sparsity arises from the compact support of the Whitney one-form basis functions and the localized interactions in the finite-element mesh. Fig.~6 illustrates the sparsity patterns of these operator-agnostic matrices from our experiments, and Table~II presents their sparsity ratios. As the degrees of freedom increase, the sparsity ratios increase because the matrices have a banded structure with fixed number of non-zero elements per row dictated by the physics and geometry of the problem.

\subsection{Efficient Computation of Refinement Kernel Matrices Using Givens Rotation QR Factorization}

In Section~III-B, we discussed the efficient computation of operator-agnostic refinement matrices \(\{\tilde{\mathbf{C}}^j\}_{j=1}^{q-1}\). For our sparse implementation, it is essential to compute the refinement kernel matrices (null spaces of the operator-agnostic refinement matrices) \(\{\tilde{\mathbf{W}}^j\}_{j=1}^{q-1}\) as defined in (16). These matrices must be obtained with high accuracy and in nearly \( \mathcal{O}(N) \) operations. Considering the banded sparse structure of the $\tilde{\mathbf{C}}^j$ matrices, we employ Givens rotation-based QR factorization to efficiently compute operator-agnostic refinement kernels.

To compute $\tilde{\mathbf{W}}^j$, the QR decomposition by Givens rotation is performed on $\tilde{\mathbf{C}}^{j,T}$, an $n_{j+1} \times n_j$ full-rank matrix. The QR decomposition yields:
\begin{equation}
    \tilde{\mathbf{C}}^{j,T} = \mathbf{Q} \mathbf{R},
\end{equation}
where $\mathbf{Q}$ is an $n_{j+1} \times n_{j+1}$ orthogonal matrix and $\mathbf{R}$ is an $n_{j+1} \times n_j$ upper triangular matrix.

Since the rank of $\tilde{\mathbf{C}}^{j,T}$ is $n_j$, the first $n_j$ columns of $\mathbf{Q}$ span the range of $\tilde{\mathbf{C}}^{j,T}$, and the remaining $n_{j+1} - n_j$ columns span its null space. Thus, $\tilde{\mathbf{W}}^j$ is obtained by taking the transpose of the last $n_{j+1} - n_j$ columns of $\mathbf{Q}$:
\begin{equation}
    \tilde{\mathbf{W}}^j = \left[ \mathbf{q}_{n_j+1}, \mathbf{q}_{n_j+2}, \dots, \mathbf{q}_{n_{j+1}} \right]^T,
\end{equation}
where $\mathbf{q}_k$ denotes the $k$-th column of $\mathbf{Q}$. This results in $\tilde{\mathbf{W}}^j$ being an $(n_{j+1} - n_j) \times n_{j+1}$ = $N_j \times n_{j+1}$ matrix whose rows form a basis for the null space of $\tilde{\mathbf{C}}^j$.

The theoretical computational complexity of QR factorization by Givens rotation is \(\mathcal{O}(N d^2)\), where \(d\) is the bandwidth of the banded matrix. In our case, since \(d \ll N\) for the matrices \(\tilde{\mathbf{C}}^j\), the computational complexity approaches \(\mathcal{O}(N)\) \cite{lindner2016recycling, golub2013matrix, saad2003iterative}. As detailed in the Section~V, our numerical experiments confirm this estimate, with observed complexities ranging between \(\mathcal{O}(N^{0.9})\) and \(\mathcal{O}(N^{1.2})\). The banded structure further limits undesired fill-ins, preserving the sparsity and improving computational efficiency. The application of appropriate threshold values in the QR decomposition algorithm ensures high accuracy with a computational cost of nearly \(\mathcal{O}(N)\). In addition, each Givens rotation modifies only two specific rows or columns of a matrix at a time. This allows computations on different portions of the matrix to be performed independently, which facilitates parallel implementations~\cite{lindner2016recycling, golub2013matrix, saad2003iterative, ipsen1984parallel}.

\subsection{Sparse Formulation}

As explained in Section~III-C, the proposed multiscale FEM approach can be reformulated to achieve nearly linear computational complexity.  Using the hierarchical structure of the algorithm and employing sparse linear algebra tools, the method can be reformulated in terms of the sparse operator-agnostic matrices $\tilde{\mathbf{C}}^j$, $\tilde{\mathbf{W}}^j$, $\tilde{\mathbf{A}}^q$, $\tilde{\mathbf{\Phi}}^q$, and vectors.
The main objective is to compute the solution given in (21). To achieve this, we need to determine the coefficients ${\mathbbm{v}}^1$ and ${\mathbbm{w}}^j$, where ${\mathbb{A}}^1 {\mathbbm{v}}^1 = {\mathbbm{g}}^1$ and ${\mathbb{B}}^j {\mathbbm{w}}^j = {\mathbbm{b}}^j$. Since vectors ${\mathbbm{g}}^j$ and ${\mathbbm{g}}^{j+1}$ are related recursively through ${\mathbbm{g}}^{j} = {\mathbb{C}}^{j} {\mathbbm{g}}^{j+1}$, ${\mathbbm{g}}^1$ can be expressed in terms of linear operators \( \{ {\mathbb{C}}^j \}_{j=1}^{q-1}\) and vector ${\mathbbm{g}}^{q}$. Similarly, ${\mathbbm{b}}^{j}$ can also be constructed hierarchically. Vectors ${\mathbbm{g}}^1$ and ${\mathbbm{b}}^{j}$ can be expressed as follows:
\begin{equation}
\begin{aligned}
    {\mathbbm{g}}^1 &= {\mathbb{C}}^1 {\mathbb{C}}^2 \cdots {\mathbb{C}}^{q-1} {\mathbbm{g}}^q, \\
    {\mathbbm{b}}^j &= \tilde{\mathbf{W}}^{j} {\mathbb{C}}^{j+1} \cdots {\mathbb{C}}^{q-1} {\mathbbm{g}}^q
\end{aligned}
\end{equation}
where \({\mathbbm{g}}^q=\tilde{\mathbf{g}}^q\). \( \{ {\mathbb{C}}^j \}_{j=1}^{q-1}\) can be efficiently expressed using operator-agnostic sparse matrices and vectors through a hierarchical approach. For instance, ${\mathbb{C}}^{q-1}$ can expressed in terms of operator-agnostic matrices and vectors as
\begin{equation}
\begin{aligned}
{\mathbb{C}}^{q-1} =
&\left (\tilde{\mathbf{C}}^{q-1} \tilde{\mathbf{C}}^{q-1,T}\right)^{-1} 
\tilde{\mathbf{C}}^{q-1} \Bigg[
{\tilde{\mathbf{I}}} - \tilde{\mathbf{A}}^{q} \tilde{\mathbf{W}}^{q-1,T} \cdots \\
&  \cdots \left(
\tilde{\mathbf{W}}^{q-1} \tilde{\mathbf{A}}^{q} \tilde{\mathbf{W}}^{q-1,T}
\right)^{-1} 
\tilde{\mathbf{W}}^{q-1}
\Bigg],
\label{eq:hierarchical_C}
\end{aligned}
\end{equation}
where $\tilde{\mathbf{I}}$ represents the identity matrix. Similarly, ${\mathbb{C}}^{q-2}$ can be expressed as
\begingroup
\small
\begin{align}
{\mathbb{C}}^{q-2} = 
& \left( \tilde{\mathbf{C}}^{q-2} \tilde{\mathbf{C}}^{q-2, T} \right)^{-1}
\tilde{\mathbf{C}}^{q-2} \cdots \nonumber \\
& \cdots \bigg[
\tilde{\mathbf{I}} 
- 
\left( 
{\mathbb{C}}^{q-1} \tilde{\mathbf{A}}^{q} {\mathbb{C}}^{q-1, T}
\right)
\tilde{\mathbf{W}}^{q-2, T}
\cdots \nonumber \\
& \cdots
\left(\tilde{\mathbf{W}}^{q-2} 
\left({\mathbb{C}}^{q-1} \tilde{\mathbf{A}}^{q}{\mathbb{C}}^{q-1, T}\right) 
\tilde{\mathbf{W}}^{q-2, T} \right)^{-1}
\tilde{\mathbf{W}}^{q-2} \bigg],
\end{align}
\endgroup
where ${\mathbb{C}}^{q-1}$ is constructed using only sparse matrices and vectors, as given in (29). By substituting the expression for ${\mathbb{C}}^{q-1}$ from (29) into (30), ${\mathbb{C}}^{q-2}$ can be formulated exclusively in terms of sparse matrix-vector multiplications. Using this approach, \( \{ \mathbb{C}^j \}_{j=1}^{q-1}\) for all resolution levels can be constructed solely using sparse matrices and vectors. Consequently, the vectors in (28), viz., the coarser-level RHS ${\mathbbm{g}}^1$ and the RHS of any detail level ${\mathbbm{b}}^j$, can also be obtained via sparse matrix-vector multiplications. Additionally, it can be easily seen that, ${\mathbb{A}}^1$ and 
$\{{\mathbb{B}}^j\}_{j=1}^{q-1}$ linear operators can be expressed using only sparse operator-agnostic matrices:
\begin{equation}
\begin{aligned}
    {\mathbb{A}}^1 &= {\mathbb{C}}^1 {\mathbb{C}}^{2} \cdots {\mathbb{C}}^{q-1} \tilde{\mathbf{A}}^q {\mathbb{C}}^{q-1,T} \cdots {\mathbb{C}}^{2,T} {\mathbb{C}}^{1,T}, \\
    {\mathbb{B}}^j &= \tilde{\mathbf{W}}^j {\mathbb{C}}^{j+1} \cdots {\mathbb{C}}^{q-1} \tilde{\mathbf{A}}^q {\mathbb{C}}^{q-1,T} \cdots {\mathbb{C}}^{j+1,T} \tilde{\mathbf{W}}^{j,T}.
\end{aligned}
\end{equation}
As a result, the final coefficients in the operator-adapted multiscale FEM approach are determined by solving equations that involve only sparse matrix-vector multiplications:
\begin{equation}
{\mathbb{C}}^1 \cdots {\mathbb{C}}^{q-1} 
\tilde{\mathbf{A}}^q {\mathbb{C}}^{q-1,T} \cdots {\mathbb{C}}^{1,T} {\mathbbm{v}}^1 
= {\mathbb{C}}^1 \cdots {\mathbb{C}}^{q-1} \tilde{\mathbf{g}}^q
\end{equation}
\begin{align}
    &\left( \tilde{\mathbf{W}}^j {\mathbb{C}}^{j+1} \cdots {\mathbb{C}}^{q-1} 
    \tilde{\mathbf{A}}^q {\mathbb{C}}^{q-1,T} \cdots {\mathbb{C}}^{j+1,T} \tilde{\mathbf{W}}^{j,T} \right) {\mathbbm{w}}^j \nonumber \\
    &\hspace{1cm}= \left( \tilde{\mathbf{W}}^j {\mathbb{C}}^{j+1} \cdots {\mathbb{C}}^{q-1} \right)\tilde{\mathbf{g}}^q, 
    \quad \text{for } j = 1, 2, \dots, q-1.
\end{align}
Equations (32) and (33) can be efficiently solved using iterative solvers as detailed in Section~IV-D. Once these coefficients are determined, both the coarser and detail-level solutions can be computed using sparse matrix-vector multiplications:
\begin{equation}
   {\mathbbm{s}}^{\text{coarse}}  = \mathoutline{{\Phi}}^{1,T} {\mathbbm{v}}^1 = \tilde{{\mathbf{\Phi}}}^{q,T} {\mathbb{C}}^{q-1,T} {\mathbb{C}}^{q-2,T} \cdots {\mathbb{C}}^{1,T}  {\mathbbm{v}}^1
\end{equation}
\begin{equation}
    {\mathbbm{s}}^{\text{detail, j}} = \mathoutline{{\Psi}}^{j,T} {\mathbbm{w}}^j = \tilde{{\mathbf{\Phi}}}^{q,T} {\mathbb{C}}^{q-1,T} \cdots {\mathbb{C}}^{j+1,T} \tilde{\mathbf{W}}^{j, T}  {\mathbbm{w}}^j
\end{equation}

As shown in (29) and (30), this formulation involves matrix inversion operations. To solve \( {\mathbb{A}}^1 {\mathbbm{v}}^1 = {\mathbbm{g}}^1 \) and \( {\mathbb{B}}^j {\mathbbm{w}}^j = {\mathbbm{b}}^j \), our hierarchical formulation incorporates multiple inverse terms across different levels. The term 
\(\left( \tilde{\mathbf{W}}^{q-1} \tilde{\mathbf{A}}^q \tilde{\mathbf{W}}^{q-1,T} \right)^{-1}\) can be used as a representative example. Instead of directly computing this matrix inverse, it is reformulated as
\begin{equation}
\tilde{\mathbf{W}}^{q-1} \tilde{\mathbf{A}}^q \tilde{\mathbf{W}}^{q-1,T} \mathoutline{\bm{\rho}} = \mathoutline{\bm{\zeta}}
\end{equation}
where \mathoutline{\bm{\zeta}} is a known vector obtained from the preceding sparse matrix-vector multiplication steps. These linear equations are solved iteratively using methods detailed in the next subsection. The complete algorithm featuring nearly \(\mathcal{O}(N)\) computational complexity is presented in Algorithm~1.

\subsection{Hierarchical Iterative Solvers}

The sparse linear systems obtained in Section~IV-C can be efficiently solved using Krylov subspace iterative solvers. In the numerical experiments considered in this work, the stiffness matrices are indefinite and non-Hermitian. Therefore, preconditioned GMRES and/or LGMRES solvers are used. When combined with the preconditioners described below, unrestarted GMRES solver achieve convergence within 50-250 iterations for a tolerance of \(\varepsilon=10^{-6}\), even for systems with millions of degrees of freedom. Similar convergence speeds are consistently observed across systems of varying sizes and scale levels.

We employ incomplete LU (ILU) preconditioners in our algorithm. However, as previously discussed, directly ILU factorization on intermediate operator-adapted matrices is not feasible because of their dense structure. To maintain the \( \mathcal{O}(N) \) computational cost, we construct the preconditioners using mimics of these intermediate matrices generated by a Sparse Approximate Inverse (SPAI)~\cite{kim2011}. In this way, we ensure that only sparse matrices are used throughout the algorithm. We denote the resulting matrices as $\{\widehat{\mathbf{A}}^j\}_{j=1}^{q}$ 
and $\{\widehat{\mathbf{B}}^j\}_{j=1}^{q-1}$. The main steps are as follows:

\begin{enumerate}
    \item \emph{Compute sparse approximate inverses of $\widehat{\mathbf{B}}^j$ matrices using SPAI:} For $j = 1, 2, \dots, q-1$, compute sparse approximate inverses of $\widehat{\mathbf{B}}^j$ using the SPAI algorithm with SPAI parameter $\kappa = 2$ or $3$~\cite{kim2011}.
    
    \item \emph{Update operator-adapted matrices:} Using the sparse inverse approximations of detail-level stiffness matrices in the matrix-based algorithm, sparse operator-adapted matrices $\widehat{\mathbf{A}}^j$ and $\widehat{\mathbf{B}}^j$ for each scale level will be obtained.

    \item \emph{Construct ILU Preconditioners:} Factorize the sparse matrices $\widehat{\mathbf{A}}^1$ and $\widehat{\mathbf{B}}^j$ to build ILU preconditioners.

    \item \emph{Solve the linear equations with GMRES/LGMRES}: After constructing the ILU preconditioners, the linear systems can be solved using preconditioned unrestarted GMRES and/or LGMRES. In our experiments, convergence was achieved in fewer than 100 iterations for a residual threshold of $\varepsilon = 10^{-6}$.
\end{enumerate}

For smaller problems (with less than about 5000 unknowns), we can directly construct sparse approximations of $\{{\mathbb{A}}^j\}_{j=1}^{q}$ and $\{{\mathbb{B}}^j\}_{j=1}^{q-1}$ = $\left\{\widetilde{\mathbf{W}}^{j}\,{\mathbb{A}}^{j+1}\,\widetilde{\mathbf{W}}^{j,T}\right\}_{j=1}^{q-1}$ using the usual FEM and the matrix-based operator-adapted wavelet decomposition algorithm. This approach yields operator-agnostic matrices suitable for ILU factorization, denoted as $\{\tilde{\mathbf{A}}^j\}_{j=1}^{q}$ and $\{\widehat{\mathbf{B}}^j\}_{j=1}^{q-1}$:
\begin{enumerate}
    \item \emph{Compute $\tilde{\mathbf{A}}^j$ with usual FEM:} Create each level's operator-agnostic stiffness matrices via usual FEM, $\tilde{\mathbf{A}}^j$.
    \item \emph{Compute $\widehat{\mathbf{B}}^j$:} For $j = 1, 2, \dots, q-1$, calculate
    \[
      \widehat{\mathbf{B}}^j 
      \;=\; 
      \tilde{\mathbf{W}}^{j} \,\tilde{\mathbf{A}}^{j+1}\, \tilde{\mathbf{W}}^{j,T}.
    \]
    \item \emph{Construct ILU Preconditioners:} Factorize the sparse matrices $\tilde{\mathbf{A}}^1$ and $\widehat{\mathbf{B}}^j$ to build ILU preconditioners.
    \item \emph{Solve the linear equations with GMRES/LGMRES}: After constructing the ILU preconditioners, the linear systems can be solved using preconditioned unrestarted GMRES and/or LGMRES. In our experiments, convergence was achieved within 200--250 iterations for a residual threshold of $\varepsilon = 10^{-6}$.
\end{enumerate}

By combining Algorithm~1 with the methods presented in Sections~III and~IV, it is possible to obtain operator-adapted wavelet decomposition-based FEM solutions with high accuracy and nearly \(\mathcal{O}(N)\) computational complexity per iteration, as verified in the numerical experiments that follow. Finally, Appendix~A provides a flowchart summarizing the proposed algorithm in Fig.~10.

\algrenewcommand\algorithmicrequire{\textbf{Inputs:}}
\algrenewcommand\algorithmicensure{\textbf{Outputs:}}

\begin{algorithm}
\caption{Matrix-Free Operator-Adapted Wavelet \newline Decomposition-Based FEM Algorithm}
\begin{algorithmic}

\Require Operator-agnostic refinement matrices \(\{\tilde{\mathbf{C}}^j\}_{j=1}^{q-1}\), 
null spaces of operator-agnostic refinement matrices (refinement kernel matrices) \(\{\tilde{\mathbf{W}}^j\}_{j=1}^{q-1}\), finest-level usual FEM stiffness matrix \(\tilde{\mathbf{A}}^q = {\mathbb{A}}^q\), finest-level usual FEM basis functions (Whitney one-forms) matrix \(\tilde{\bm{\Phi}}^q = \mathoutline{\Phi}^q\), finest-level usual FEM right-hand side vector \(\tilde{\mathbf{g}}^q = {\mathbbm{g}}^q\)

\vspace{0.5em}
\noindent \textbf{\textit{Create \(\{{\mathbb{C}}^j\}_{j=1}^{q-1}\) and  \(\{{\mathbb{A}}^j\}_{j=1}^{q}\) linear operators:}}
\State Construct linear operator dictionaries for sparse matrix-vector multiplications:
\For{\(j = q\) \textbf{down to} \(2\)}
\Statex
\State \textit{${\mathbb{C}}^{j-1}$ linear operator:}
\[
\begin{aligned}
{\mathbb{C}}^{j-1} =
&\left (\tilde{\mathbf{C}}^{j-1} \tilde{\mathbf{C}}^{j-1,T}\right)^{-1} 
\tilde{\mathbf{C}}^{j-1} \Bigg[
\tilde{\mathbf{I}} - {\mathbb{A}}^{j} \tilde{\mathbf{W}}^{j-1,T} \cdots \\
&  \cdots \left(
\tilde{\mathbf{W}}^{j-1} {\mathbb{A}}^{j} \tilde{\mathbf{W}}^{j-1,T}
\right)^{-1} 
\tilde{\mathbf{W}}^{j-1}
\Bigg]
\end{aligned}
\]
\State \textit{${\mathbb{A}}^{j-1}$ linear operator:}
\[
{\mathbb{A}}^{j-1} = {\mathbb{C}}^{j-1} {\mathbb{A}}^j {\mathbb{C}}^{j-1,T}
\]
\EndFor

\vspace{0.5em}
\noindent {{${\mathbb{C}}^{j-1}$ and ${\mathbb{A}}^{j-1}$ are linear operators, defined hierarchically using only operator-agnostic sparse matrices and vectors.}}

\vspace{0.5em}
\noindent \textbf{\textit{Coarsest level right-hand side can be calculated as}:}
\[
{\mathbbm{g}}^1 = {\mathbb{C}}^1 {\mathbb{C}}^2 \dots {\mathbb{C}}^{q-1} \tilde{\mathbf{g}}^q
\]

\State Solve the linear system
\[
{\mathbb{A}}^1 {\mathbbm{v}}^1 = {\mathbbm{g}}^1
\]

\vspace{0.5em}
\noindent {{via preconditioned GMRES/LGMRES iterative solvers, as discussed in Section~IV-D}.}

\vspace{0.5em}

\noindent \textbf{\textit{Coarsest level solution}:}

\[
{\mathbbm{s}}^{\text{coarse}} = \tilde{\mathbf{\Phi}}^{q,T}{\mathbb{C}}^{q-1,T} \cdots {\mathbb{C}}^{1,T} {\mathbbm{v}}^1
\]
\vspace{0.5em}
\noindent \textbf{\textit{Detail Level Solutions:}}

\State Compute detail level coefficients and solutions
\newline
(\(\zeta\) is selected based on the desired accuracy):

\For{\(j = 2\) \textbf{to} \(\zeta\)} 
\State The detail-level coefficients are computed as:
\[
\tilde{\mathbf{W}}^{j-1} {\mathbb{A}}^j \tilde{\mathbf{W}}^{j-1,T} {\mathbbm{{w}}}^{j-1} = 
\tilde{\mathbf{W}}^{j-1} {\mathbb{C}}^{j}{\mathbb{C}}^{j+1} \dots {\mathbb{C}}^{q-1} \tilde{\mathbf{g}}^q
\]
\State The detail-level solutions are computed as:
\[
{\mathbbm{s}}^{\text{detail, j-1}} = \tilde{\mathbf{\Phi}}^{q,T}{\mathbb{C}}^{q-1,T} \cdots {\mathbb{C}}^{j,T}\tilde{\mathbf{W}}^{j-1,T} {\mathbbm{w}}^{j-1}
\]
\EndFor

\vspace{0.5em}
\noindent \textbf{\textit{Final Solution Assembly:}}
\[
\mathbbm{u}^\zeta ={\mathbbm{s}}^{\text{coarse}} + \sum_{j=1}^{\zeta-1} {\mathbbm{s}}^{\text{detail, j}}
\]

\end{algorithmic}
\end{algorithm}

\section{Numerical Experiments\label{sec:res}}

In this section, the computational efficiency and accuracy of the proposed operator-adapted wavelet decomposition-based FEM algorithm are demonstrated through its application to several representative two-dimensional multiscale electromagnetic problems: L- and U-shaped waveguide discontinuities, and a leaky waveguide comprised of two parallel MPSi slabs.

\subsection{Accuracy}

\subsubsection{L- and U-Shaped Waveguide Discontinuities}
We first investigate L- and U-shaped waveguide discontinuities that exhibit multiscale features such as sharp corners using the proposed algorithm. These experiments highlight key algorithmic concepts, including coarser and detail-level representations, and the advantages of operator orthogonality. The waveguides are excited at the left (input) port in the fundamental $\mathrm{TE}_{10}$ mode. Near the waveguide corners, the electric field comprises a superposition of the fundamental mode and rapidly decaying higher-order evanescent modes. Part of the energy is reflected back into the input port, while the remainder is transmitted to the right (output) port. Except for the input and output ports, all boundaries are modeled as perfect electric conductors. During refinement, since the geometric information of the boundary entities (e.g., coordinates and connectivity) is available, the corresponding boundary edges are identified and the same boundary conditions are assigned consistently at each resolution level. The elemental terms of the FEM stiffness matrix, $[K^e]$, and of the excitation vector, $\{b^e\}$, for this problem are given by
\begin{align}
[K^e] = & \int_{\Omega} (\nabla \times \mathbf{W}^e) \cdot (\nabla \times \mathbf{W}^e) \, d\Omega \nonumber \\
& - \beta_{10}^2 \int_{\Omega} \mathbf{W}^e \cdot \mathbf{W}^e \, d\Omega \nonumber \\
& + j \beta_{10} \int_{\Sigma, \Gamma_p^{i}} (\hat{\mathbf{n}} \times \mathbf{W}^e) \cdot (\hat{\mathbf{n}} \times \mathbf{W}^e) \, d\Gamma
\end{align}

\begin{equation}
\{b^e\} = 2j \beta_{10} \int_{\Gamma_p^{\text{input}}} (\hat{\mathbf{n}} \times \mathbf{W}^e) \cdot (\hat{\mathbf{n}} \times \mathbf{E}_{\text{inc}}) \, d\Gamma
\end{equation}

\begin{figure}[h]
    \centering

    
    \begin{subfigure}[t]{0.48\textwidth}
        \centering
        \includegraphics[width=\linewidth]{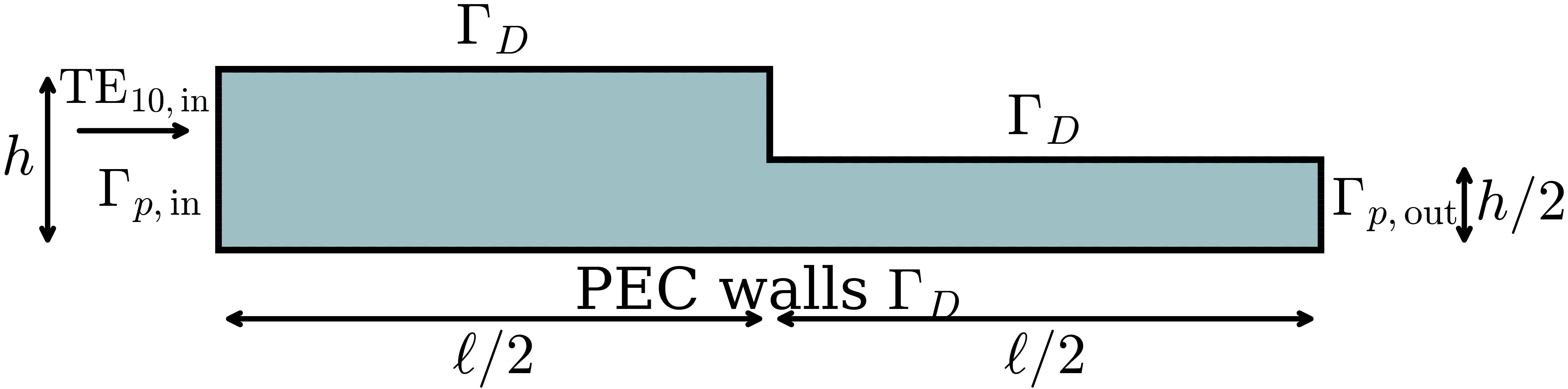}
        \caption{}
        \label{fig:wg_config_L}
    \end{subfigure}

    \vspace{0.3cm}
    \begin{subfigure}[b]{0.45\textwidth}
        \centering
        \includegraphics[width=\textwidth]{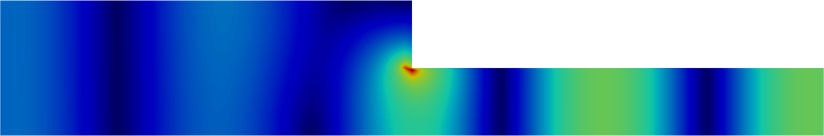} 
        \caption{}
        \label{fig:2a}
    \end{subfigure}
    
    \vspace{0.3cm}
    \begin{subfigure}[b]{0.45\textwidth}
        \centering
        \includegraphics[width=\textwidth]{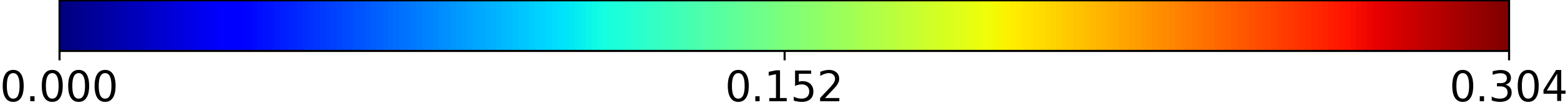} 
    \end{subfigure}
    
    \vspace{0.8cm} 
    
    \begin{subfigure}[b]{0.45\textwidth}
        \centering
        \includegraphics[width=\textwidth]{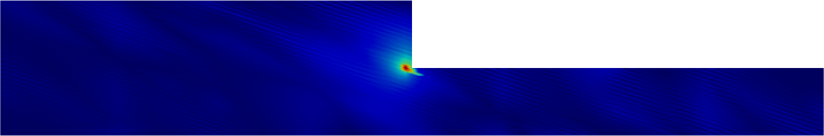} 
        \caption{}
        \label{fig:2b}
    \end{subfigure}
    
    \vspace{0.3cm}
    \begin{subfigure}[b]{0.45\textwidth}
        \centering
        \includegraphics[width=\textwidth]{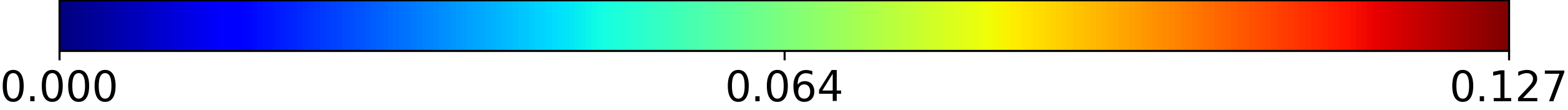} 
    \end{subfigure}
    \vspace{0.8cm} 

    \begin{subfigure}[b]{0.45\textwidth}
        \centering
        \includegraphics[width=\textwidth]{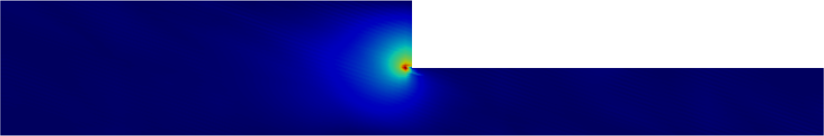} 
        \caption{}
        \label{fig:2c}
    \end{subfigure}
    
    \vspace{0.3cm}
    \begin{subfigure}[b]{0.45\textwidth}
        \centering
        \includegraphics[width=\textwidth]{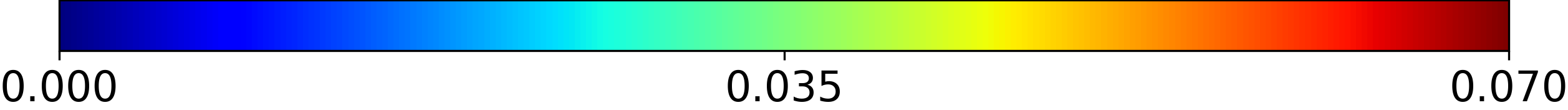} 
    \end{subfigure}
    
    \caption{L-shaped waveguide discontinuity numerical example analyzed with the sparse operator-adapted wavelet decomposition-based FEM. (a) Problem configuration. (b) Computed electric-field magnitude of the two-level solution, $\mathbbm{u}=\mathbbm{s}^{\text{coarse}}+\mathbbm{s}^{\text{detail,1}}$. (c) Second detail-level contribution, $\mathbbm{s}^{\text{detail,2}}$. (d) Third detail-level contribution, $\mathbbm{s}^{\text{detail,3}}$.}

\end{figure}

Here, $\hat{\mathbf{n}}$ denotes the outward unit normal to the boundary $\Gamma$, $\beta_{10}$ is the propagation constant of the $\mathrm{TE}_{10}$ mode, $\mathbf{W}^e$ represents the edge element basis functions, and $\mathbf{E}_{\text{inc}}$ is the incident $\mathrm{TE}_{10}$ field at the input port. The global system is assembled and solved using $K{u} = {b}$. The L- and U-shaped waveguide discontinuity configurations considered in this paper are shown in Figs.~2(a) and~3(a), respectively. Further mathematical analysis and detailed descriptions of these waveguide discontinuity problems are available in~\cite{garcia2007two}.

The maximum edge length \(\ell_{\max}\) of the triangular mesh elements is set to \(\lambda/10\) for the coarsest level and is halved at each subsequent resolution level. Fig.~2(b) illustrates a two-level solution composed of the coarsest level and the first detail level, represented as \({\mathbbm{u}} = {\mathbbm{s}}^{\text{coarse}} + {\mathbbm{s}}^{\text{detail, 1}}\). When higher accuracy is required for the overall solution, the next detail level, \({\mathbbm{s}}^{\text{detail, 2}}\), shown in Fig.~2(c), is incorporated into the solution, yielding a three-level solution. Fig.~2(d) illustrates the third detail level, \({\mathbbm{s}}^{\text{detail, 3}}\).

As illustrated in Fig.~2, both \(\mathbbm{s}^{\text{detail,2}}\) and
\(\mathbbm{s}^{\text{detail,3}}\) have small magnitudes in smooth regions yet exhibit significant concentration around the sharp corners. 
The magnitude of the detail-level solutions inherently serves as an indicator the accuracy of the overall solution. If the magnitudes of a detail level fall below a certain threshold, the subsequent detail levels do not need to be included. 

In this example, when the total solution is calculated as \(\mathbbm{u} = \mathbbm{s}^{\text{coarse}} + \mathbbm{s}^{\text{detail,1}} + \mathbbm{s}^{\text{detail,2}} + \mathbbm{s}^{\text{detail,3}}\), the result agrees with the usual FEM solution at the finest level, exhibiting a relative \(L^2\)-norm error on the order of user-defined thresholds. For instance, repeated simulations show that \(\|\tilde{\mathbf{u}}^{\text{FEM}} - \mathbbm{u}^{\text{Algorithm}}\| / \|\tilde{\mathbf{u}}^{\text{FEM}}\| \approx 2.5 \times 10^{-6}\), where \(\tilde{\mathbf{u}}^{\text{FEM}}\) is the finest-level usual FEM solution, and \(\mathbbm{u}^{\text{Algorithm}}\) is produced by Algorithm~1. This accuracy is further verified through comparisons with the numerical mode-matching (NMM) method, yielding \(\|\mathbf{u}^{\text{NMM}} - \mathbbm{u}^{\text{Algorithm}}\| / \|\mathbf{u}^{\text{NMM}}\| = 1.21 \times 10^{-4}\), where \(\mathbf{u}^{\text{NMM}}\) denotes the NMM solution.

\begin{table}[h]
    \centering
    \renewcommand{\arraystretch}{1.2} 
    \caption{Relative energy contents of different scale levels}
    \label{tab:energy_content}
    \begin{tabular}{>{\centering\arraybackslash}m{4cm} >{\centering\arraybackslash}m{4cm}}
        \toprule
        \textbf{Scale level} & \textbf{Relative energy content (\%)} \\
        \midrule
        $\mathbbm{s^{\text{coarse}}}$           & 57.2\% \\
        $\mathbbm{s^{\text{detail,1}}}$         & 28.7\% \\
        $\mathbbm{s^{\text{detail,2}}}$         & 10.88\% \\
        $\mathbbm{s^{\text{detail,3}}}$         & 3.22\% \\
        \bottomrule
    \end{tabular}
\end{table}

Table~I presents the relative energy content (in percentage) for each scale in the $L$-shaped waveguide scenario. As previously discussed, the coarser-level solution is operator-orthogonal to each detail-level solution, and all detail-level solutions are operator-orthogonal to one another. Consequently, the finest-level solution is the direct sum of the coarser-level and all subsequent detail-level solutions under the operator \( \mathcal{L} \), as given in (13). As expected, due to the inherent nature of the method, the coarse level and the first few detail levels capture most of the energy content.

The accuracy of the proposed algorithm is also evaluated using a U-shaped waveguide discontinuity problem, as shown in Fig.~3, where the effects of the discontinuities are particularly prominent. Fig.~3(b) illustrates the \({\mathbbm{u}} = {\mathbbm{s}}^{\text{coarse}} + {\mathbbm{s}}^{\text{detail, 1}}\) solution for this problem. As shown, the electric field magnitude again exhibits dominant behavior around the sharp corners due to higher-order evanescent modes, while significantly lower in smooth regions. 
In this example, the maximum edge length \(\ell_{\max}\) of the triangular elements is set to \(\lambda/12.5\) for the coarsest level and is halved at each subsequent resolution level. When a five-scale solution is computed, the relative \(L^2\)-norm error between the solution of the finest-level usual FEM and the solution obtained using the proposed algorithm is approximately \(8 \times 10^{-6}\). Furthermore, comparing the solution obtained from the proposed algorithm with that of the numerical mode-matching method yields a \(L^2\)-norm error of \(7.6 \times 10^{-4}\).

\begin{figure}[h]
    \centering


    \begin{subfigure}[t]{0.48\textwidth}
        \centering
        \includegraphics[width=\linewidth]{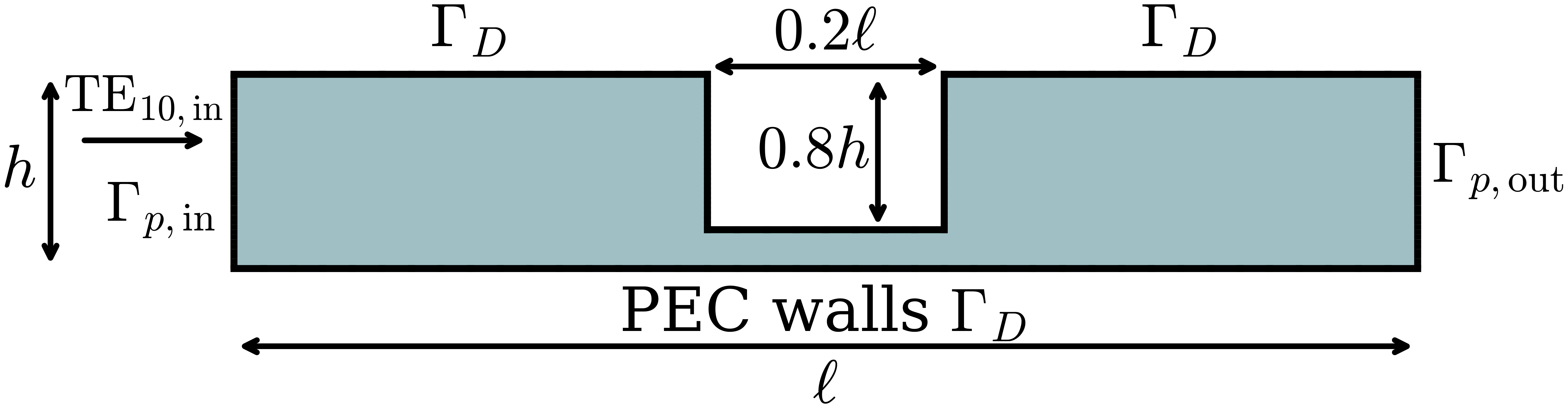}
        \caption{}
        \label{fig:wg_config_U}
    \end{subfigure}
    
    \begin{subfigure}[b]{0.45\textwidth}
        \centering
        \includegraphics[width=\textwidth]{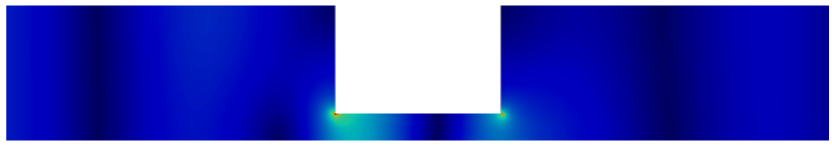} 
        \caption{}
        \label{fig:3a}
    \end{subfigure}
    
    \vspace{0.3cm}
    \begin{subfigure}[b]{0.45\textwidth}
        \centering
        \includegraphics[width=\textwidth]{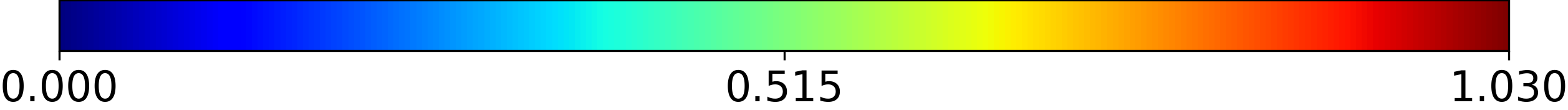} 
    \end{subfigure}
    
    \vspace{0.8cm} 
    
    \begin{subfigure}[b]{0.45\textwidth}
        \centering
        \includegraphics[width=\textwidth]{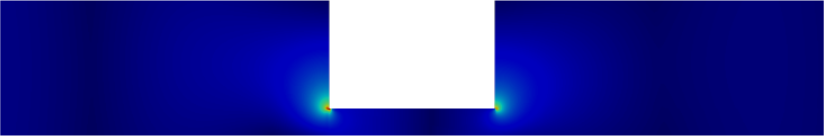} 
        \caption{}
        \label{fig:3b}
    \end{subfigure}
    
    \vspace{0.3cm}
    \begin{subfigure}[b]{0.45\textwidth}
        \centering
        \includegraphics[width=\textwidth]{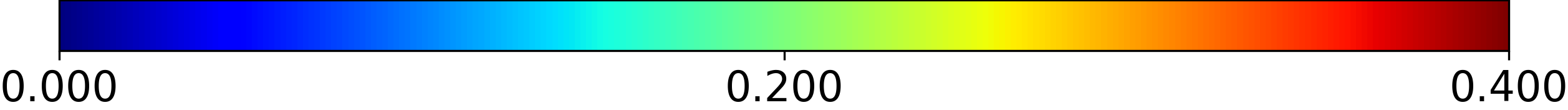} 
    \end{subfigure}
    \vspace{0.8cm} 

    \begin{subfigure}[b]{0.45\textwidth}
        \centering
        \includegraphics[width=\textwidth]{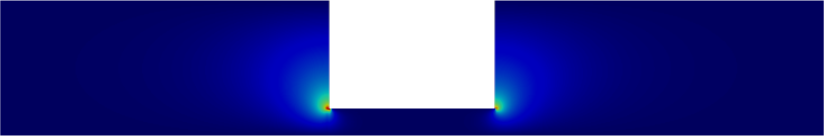} 
        \caption{}
        \label{fig:3c}
    \end{subfigure}
    
    \vspace{0.3cm}
    \begin{subfigure}[b]{0.45\textwidth}
        \centering
        \includegraphics[width=\textwidth]{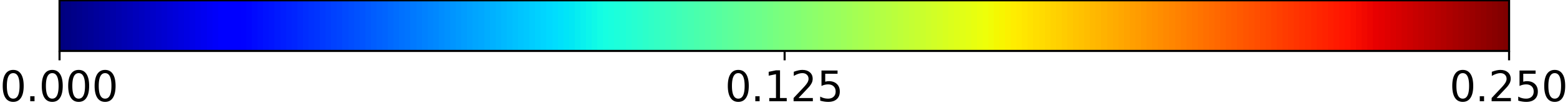} 
    \end{subfigure}

    \caption{U-shaped waveguide discontinuity numerical example analyzed with the sparse operator-adapted wavelet decomposition-based FEM. (a) Problem configuration. (b) Computed electric-field magnitude of the two-level solution, $\mathbbm{u}=\mathbbm{s}^{\text{coarse}}+\mathbbm{s}^{\text{detail,1}}$. (c) Second detail-level contribution, $\mathbbm{s}^{\text{detail,2}}$. (d) Third detail-level contribution, $\mathbbm{s}^{\text{detail,3}}$.}

\end{figure}

\subsubsection{Leaky MPSi Waveguide}

We next consider a leaky waveguide composed of two parallel MPSi slabs and excited by two in-phase point sources. This problem involves multiple geometric length scales, most notably a sub-wavelength rectangular lattice of cylindrical air pores, high-permittivity dielectric loading, and strong near-field coupling across the inter-slab air gap, making it a demanding benchmark for the proposed operator-adapted wavelet-decomposition-based FEM. The MPSi slab can also operate in a superlensing regime, where the transmitted field forms a focus with a width smaller than the classical diffraction-limited spot \cite{donderici2005subgridding,luo2002aanr}.

The computational domain is a 2-D rectangular waveguide of size $L_x \times L_y = 8\lambda_0 \times 8\lambda_0$, where $\lambda_0$ is the free-space wavelength. Two identical perforated MPSi slabs are placed parallel to each other. The host material (Si) is modeled with relative permittivity $\epsilon_{r,\mathrm{MPSi}}=11.65$, while the pores are modeled as cylindrical air holes ($\epsilon_r=1$) arranged on a rectangular lattice of period
$a = 0.1294\,\lambda_0$ in both directions. The hole diameter is
$d_h = 0.1176\,\lambda_0$ (radius $r_h = 0.0588\,\lambda_0$). Each slab has thickness
$w=6a \approx 0.7765\,\lambda_0$ (six pore columns across the thickness), with hole centers located at
$x = x_{\min} + (m+\tfrac{1}{2})a$, $m=0,1,\ldots,5$.

\begin{figure}[h]
  \centering
  \includegraphics[width=1\linewidth]{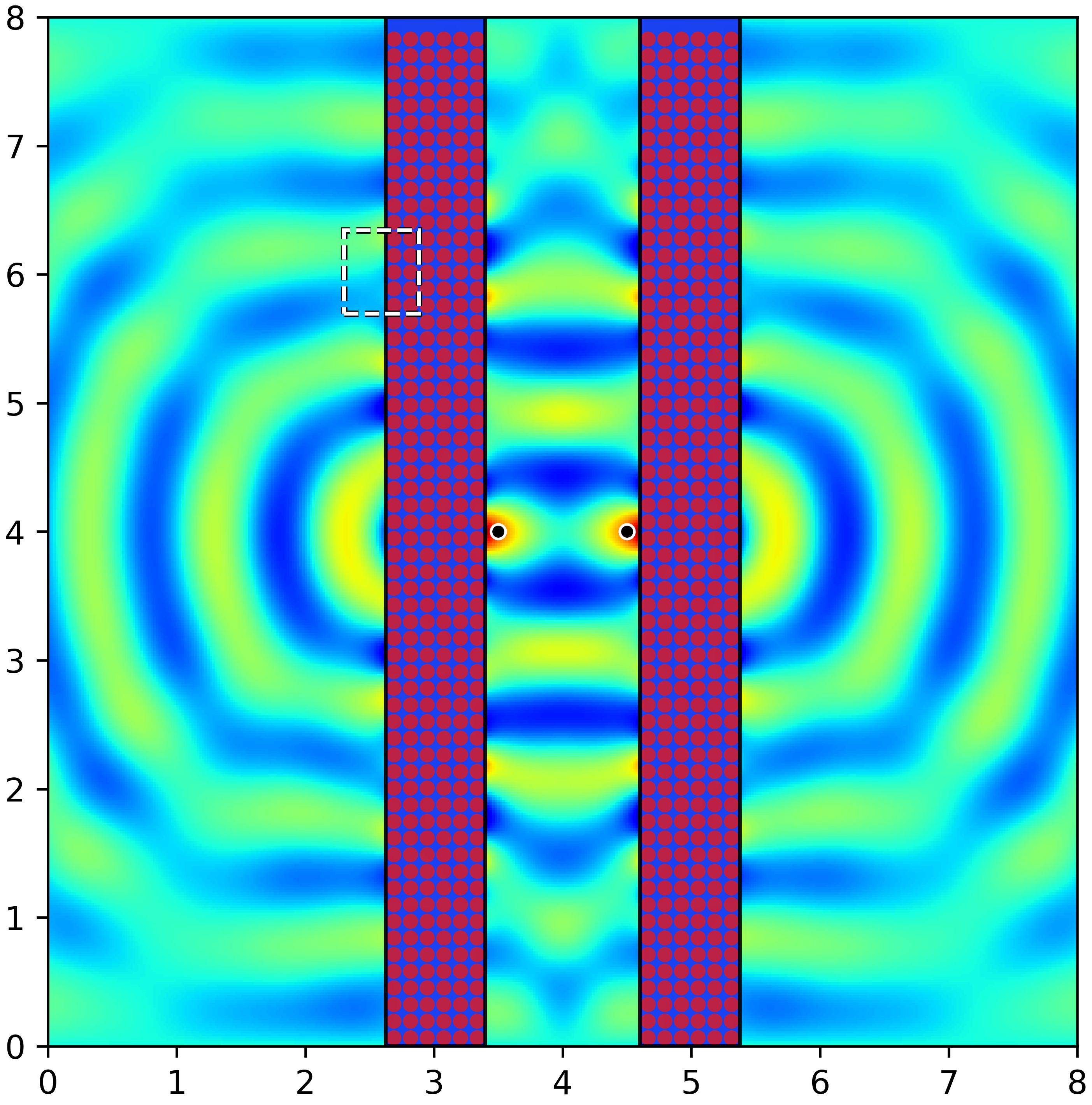}
  \captionsetup{width=1\linewidth}
  \caption{Electric field distribution in the leaky MPSi waveguide, computed with a five-level multiscale method. The total field is reconstructed as $\mathbbm{E}=\mathbbm{E}^{\text{coarse}}+\sum_{k=1}^{4}\mathbbm{E}^{\text{detail},\,k}$. The two in-phase point sources are marked by black dots, and the white dashed rectangle highlights the region whose multilevel meshes are shown in Fig.~\ref{Fig-mesh}. Axes are in units of the wavelength $\lambda$.}

\end{figure}

Fig.~4 shows the two-slab waveguide configuration. The slabs are separated by an air gap of $g=1.2\lambda_0$ and are excited by two point sources as indicated in Fig.~4; the sources are separated by $d_s=1.0\lambda_0$ and each is offset by $\delta=0.1\lambda_0$ from the nearest slab interface. Fig.~4 also illustrates the electric-field solution obtained with a five-level ($q=5$) multiscale hierarchy. When the total field is assembled by incorporating all scale contributions, i.e., $\mathbbm{u} = \mathbbm{s}^{\text{coarse}}+\sum_{j=1}^{4}\mathbbm{s}^{\text{detail},\,j}$, the proposed method agrees very closely with the usual finest-level FEM solution, yielding a relative $L^2$-norm error of approximately $2.7\times 10^{-5}$. Beyond accuracy, this configuration is particularly challenging because it combines strongly multiscale geometric features, including a sub-wavelength air-hole lattice embedded in a high-permittivity MPSi host with thin dielectric ligaments between adjacent pores, with pronounced near-field interactions in the inter-slab region. Nevertheless, the proposed approach delivers precise results. The CPU-time measurements across problems with different geometric scales confirm that the near-linear computational complexity trend persists, as discussed in Subsection~V-C.

\begin{figure}[h]
    \centering
    \begin{subfigure}{1\linewidth} 
        \centering
        \includegraphics[width=0.95\linewidth]{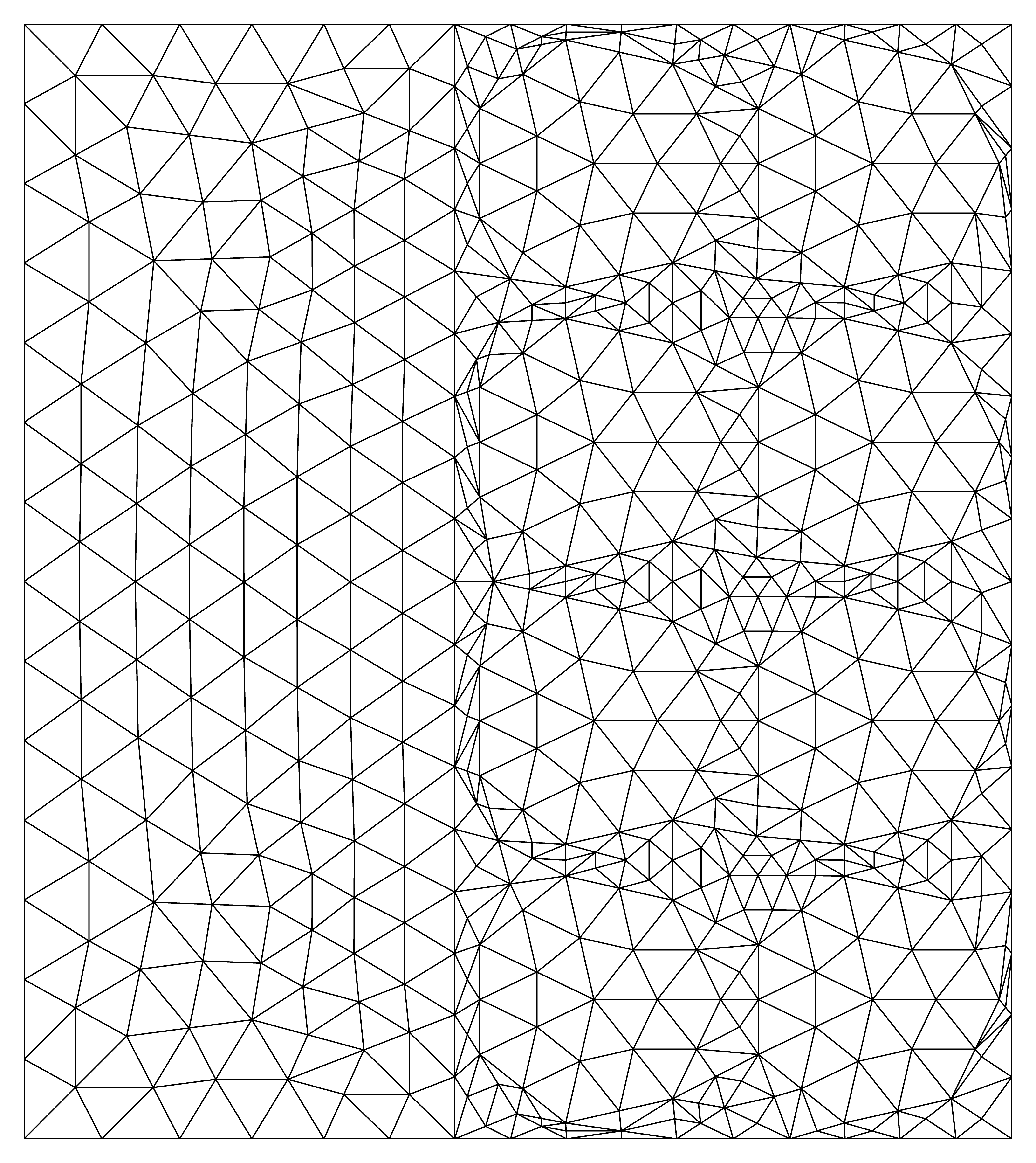} 
        \caption{}
    \end{subfigure}
    \hspace{0.1cm} 
    \begin{subfigure}{1\linewidth} 
        \centering
        \includegraphics[width=0.95\linewidth]{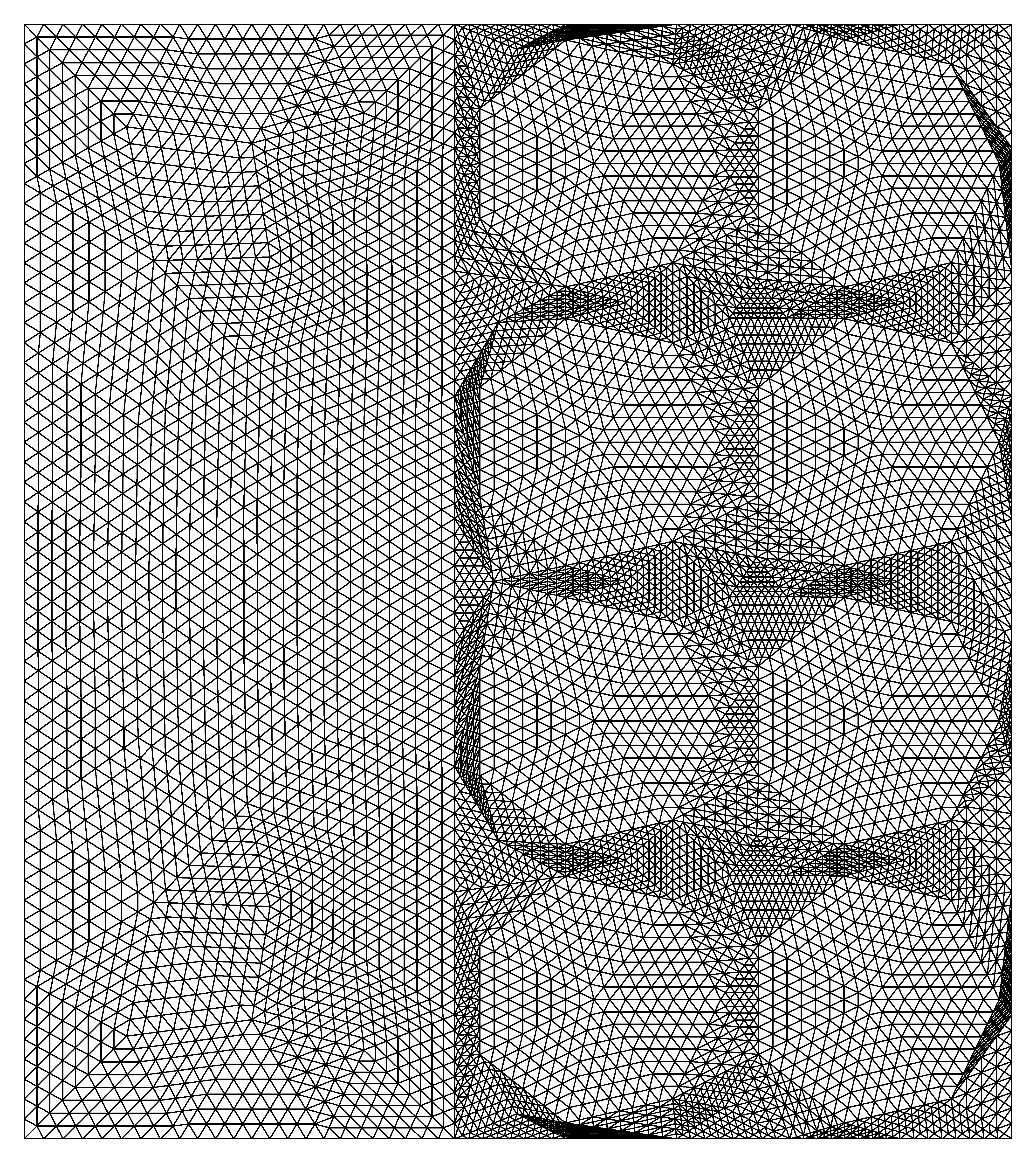} 
        \caption{}
    \end{subfigure}

    \caption{Zoomed view of the meshes corresponding to the region indicated by the white dashed rectangle in Fig.~4, for the five-level mesh hierarchy: (a) coarsest-level mesh and (b) third-level mesh.}
\label{Fig-mesh}
\end{figure}

Figs.~5(a) and~5(b) show the coarsest-level and third-level meshes, respectively, of the five-level mesh hierarchy, restricted to the region of interest highlighted by the white dashed rectangle in Fig.~4. As can be seen from figures, the proposed method can be applied with the similar level of accuracy to mesh hierarchies obtained by successively refining an adaptive coarsest-level mesh. The computational gains offered by the present framework can be further enhanced by combining it with adaptive coarsening strategies applied to finest-level adaptive meshes. In particular, constructing an adaptive multilevel hierarchy with convex polygonal elements can represent pores and planar interfaces more economically at coarser levels, thereby reducing the total number of unknowns and improving overall efficiency. While such coarsening-based polygonal hierarchies are part of our ongoing work, the results reported in this paper employ the refinement-based mesh hierarchy, and demonstrate that the sparse operator-adapted wavelet decomposition method detailed in Algorithm~1 can be applied precisely with nearly linear computational complexity.

\subsection{Sparsity}

Algorithm~1 hierarchically constructs coarse- and detail-level solutions using only sparse operator-agnostic matrices and vectors. As discussed in Section~IV-A, the computational efficiency of the method arises from the high sparsity of these operator-agnostic matrices. Fig.~6 illustrates the sparsity patterns of \(\tilde{\mathbf{C}}\), \(\tilde{\mathbf{W}}\), and \(\tilde{\mathbf{A}}\) matrices obtained from the L-shaped waveguide analysis, while Table~II presents their sparsity ratios for various DoFs. Similar sparsity patterns, with consistently high sparsity ratios, are observed across all numerical experiments considered, including the U-shaped and iris-shaped \cite{sik2025multiscale} waveguides, and the leaky MPSi waveguide. Table~II shows that the sparsity ratios of the  \(\tilde{\mathbf{C}}\), \(\tilde{\mathbf{W}}\), and \(\tilde{\mathbf{A}}\) matrices increase with DoFs due to their banded structure with fixed number of non-zero entries per row. It is important to note that, owing to the highly sparse and banded structure of these operator-agnostic input matrices and the proposed hierarchical multiscale algorithm, we observed reductions in peak memory usage of \(20\%\)–\(30\%\), depending on the number of levels and DoFs.

\begin{figure}[h]
    \centering
    \begin{subfigure}{0.4\linewidth} 
        \centering
        \includegraphics[width=\linewidth]{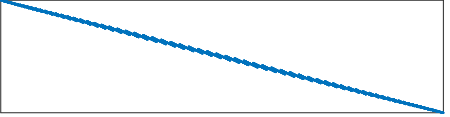} 
        \caption{}
    \end{subfigure}
    \hspace{0.1cm} 
    \begin{subfigure}{0.4\linewidth} 
        \centering
        \includegraphics[width=\linewidth]{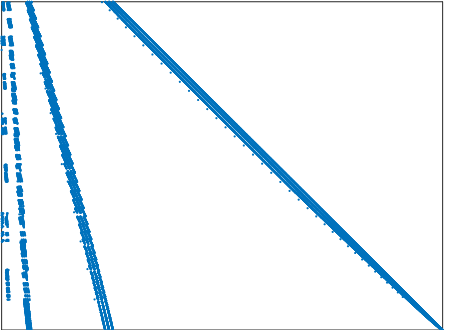} 
        \caption{}
    \end{subfigure}

    \vspace{0.15cm}

    \begin{subfigure}{0.4\linewidth} 
        \centering
        \includegraphics[width=\linewidth]{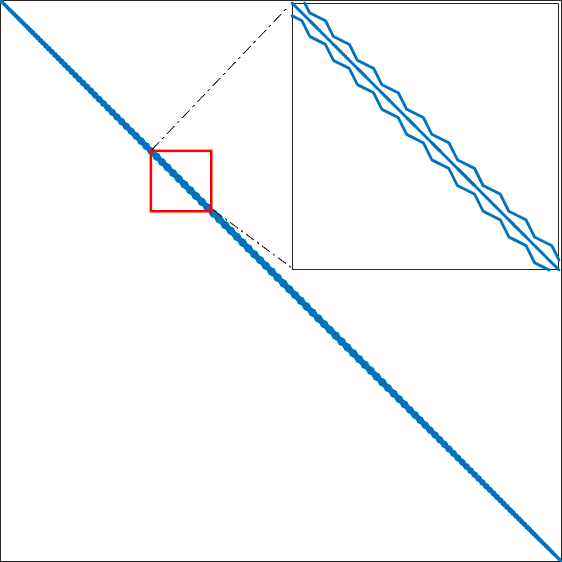} 
        \caption{}
    \end{subfigure}

    \caption{Sparsity patterns of (a) $\tilde{\mathbf{C}}^j$ (size: \(n_j \times n_{j+1}\)), (b)  $\tilde{\mathbf{W}}^j$  (size: \(N_j \times n_{j+1}\)), and (c)  $\tilde{\mathbf{A}}^j$ (size: \(n_{j+1} \times n_{j+1}\)) matrices, the red rectangle highlights the zoomed region, displayed in the inset.} 
\end{figure}

\begin{figure}[h]
    \centering
    \includegraphics[width=1\linewidth]{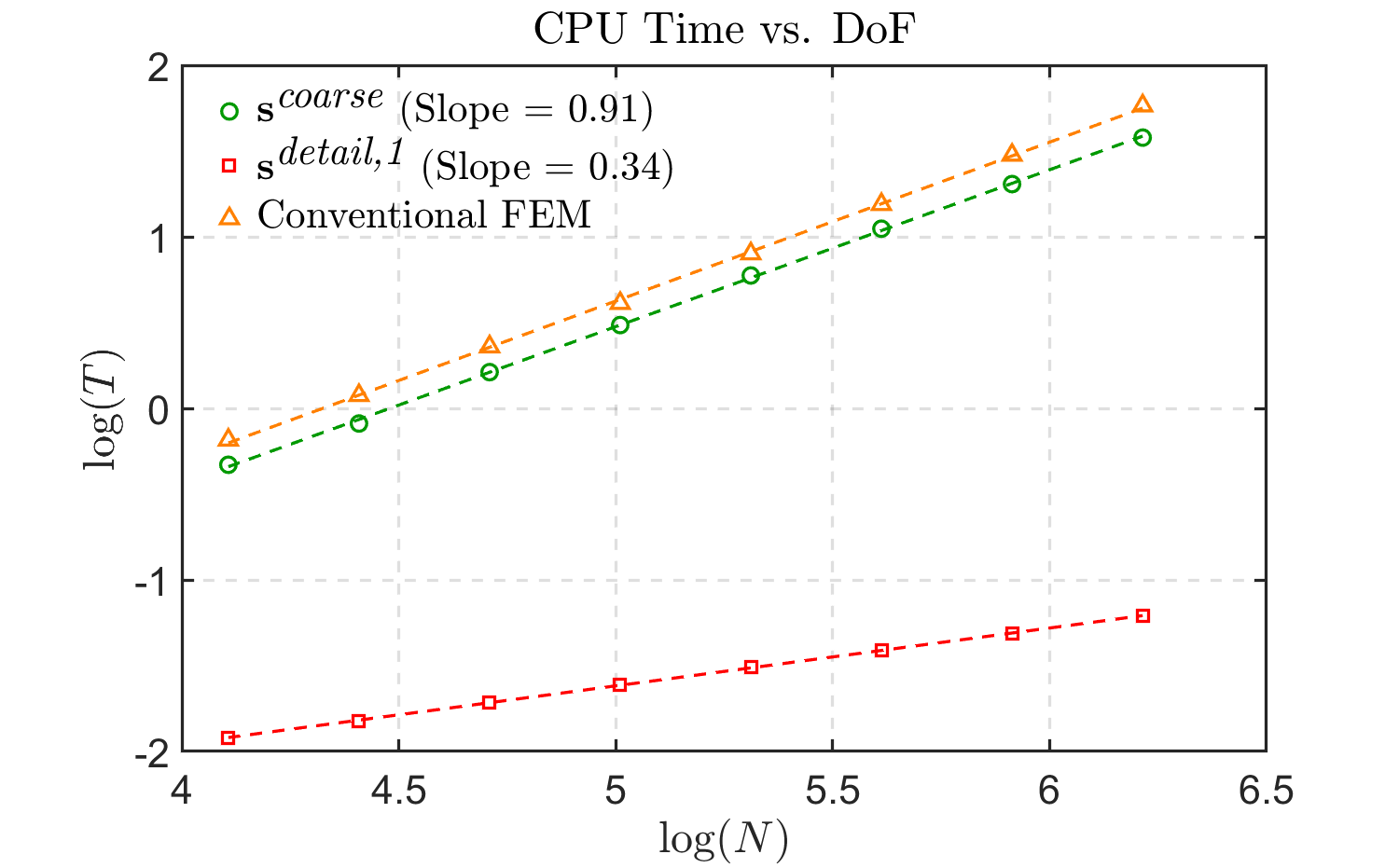}
    \captionsetup{width=1\linewidth} 
     \caption{Elapsed time per iteration versus DoF for the coarser ($\mathbbm{s^{\text{coarse}}}$) and detail ($\mathbbm{s^{\text{detail,1}}}$) level solutions for the two-level scenario, where $\mathbbm{u} = \mathbbm{s^{\text{coarse}}} + \mathbbm{s^{\text{detail,1}}}$. 
     The result for the conventional FEM solution is also provided.}
\end{figure}

\begin{figure}[h]
    \centering
    \begin{subfigure}{0.8\linewidth} 
        \centering
        \includegraphics[width=\linewidth]{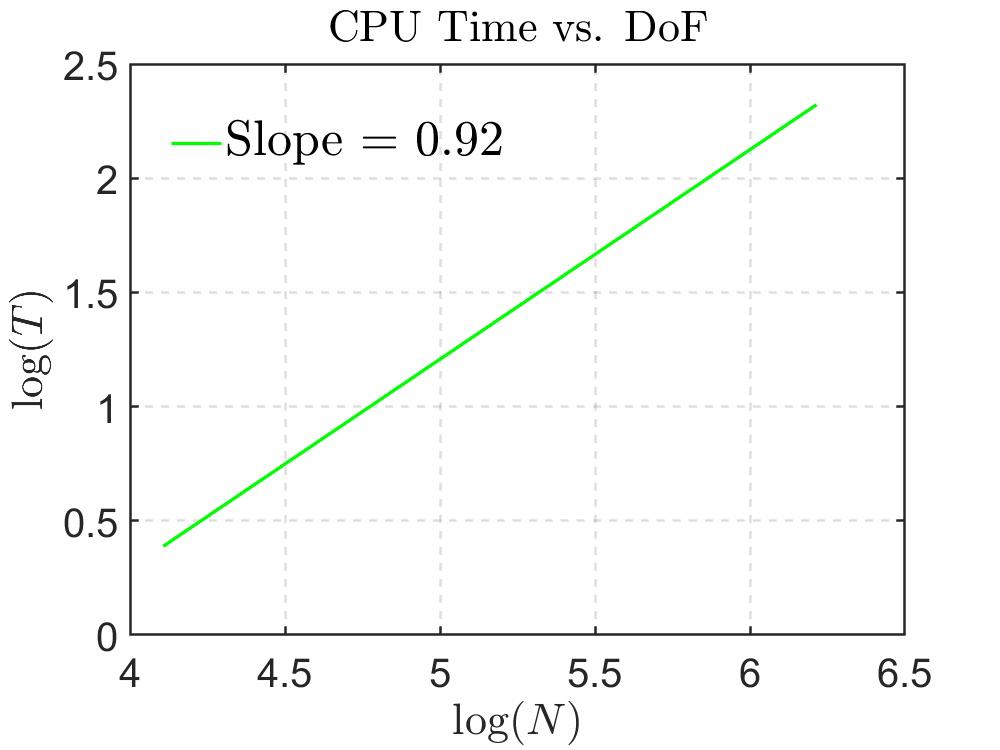}
        \caption{}
    \end{subfigure}
    \hspace{0.1cm} 
    \begin{subfigure}{0.8\linewidth} 
        \centering
        \includegraphics[width=\linewidth]{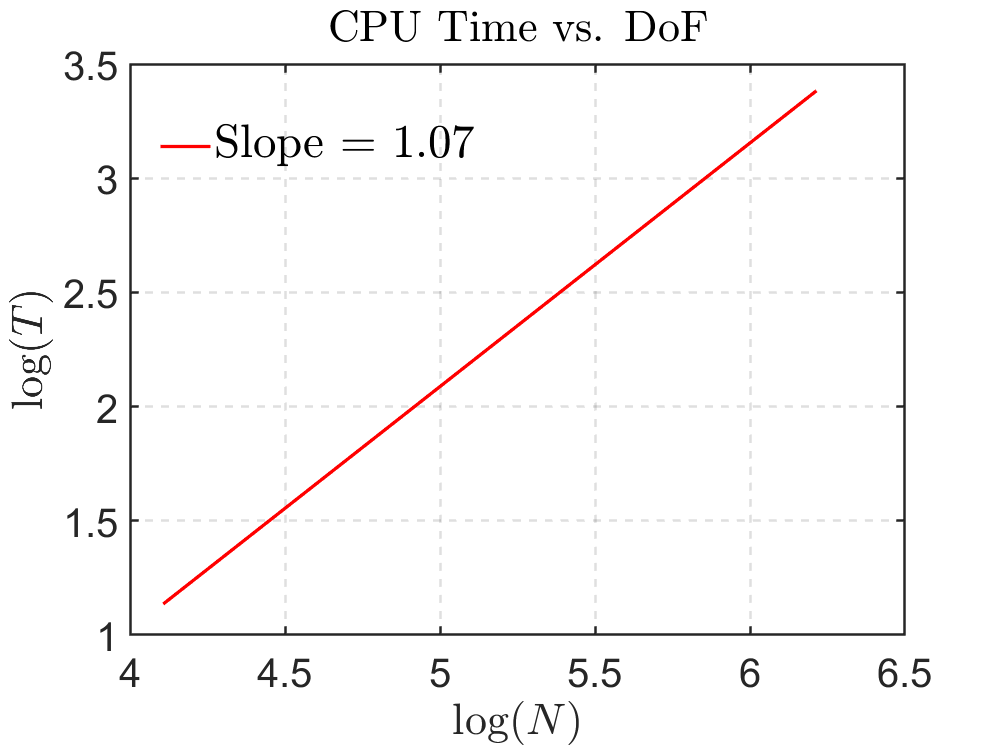} 
        \caption{}
    \end{subfigure}

    \caption{Elapsed time per iteration versus DoF for the operator-adapted wavelet decomposition FEM algorithm with six scales. (a) Result excluding precomputation steps. (b) Result including precomputation steps.} 
\end{figure}

\begin{table*}[h!]
\caption{Sparsity of $\tilde{\mathbf{C}}$, $\tilde{\mathbf{W}}$, and $\tilde{\mathbf{A}}$ matrices}
\label{tab:matrix_properties}
\centering
\small 
\renewcommand{\arraystretch}{1.2}
\setlength{\tabcolsep}{8pt}
\begin{tabular}{lcrcrcr}
\toprule
& \multicolumn{2}{c}{$\tilde{\mathbf{C}}$} & \multicolumn{2}{c}{$\tilde{\mathbf{W}}$} & \multicolumn{2}{c}{$\tilde{\mathbf{A}}$} \\[3pt]
\cmidrule(lr){2-3} \cmidrule(lr){4-5} \cmidrule(lr){6-7}
 & \textbf{Matrix Size} & \textbf{Sparsity (\%)} 
 & \textbf{Matrix Size} & \textbf{Sparsity (\%)}
 & \textbf{Matrix Size} & \textbf{Sparsity (\%)} \\
\midrule
1 & $1200 \times 4704$   & 99.8351 & $3504 \times 4704$   & 99.1543 & $4704 \times 4704$   & 99.8954 \\
2 & $4704 \times 18624$  & 99.9577 & $13920 \times 18624$ & 99.7784 & $18624 \times 18624$ & 99.9734 \\
3 & $18624 \times 74112$ & 99.9893 & $55488 \times 74112$ & 99.9433 & $74112 \times 74112$ & 99.9933 \\
4 & $74112 \times 295680$& 99.9973 & $221568 \times 295680$ & 99.9855 & $295680 \times 295680$ & 99.9983 \\
5 & $295680 \times 1181184$& 99.9992 & $885504 \times 1181184$ & 99.9963 & $1181184 \times 1181184$ & 99.9996 \\
\bottomrule
\end{tabular}
\end{table*}

\subsection{Computational Complexity}

As discussed in Section~IV, the entire algorithm, including its precomputation steps, is theoretically expected to achieve nearly \(\mathcal{O}(N)\) computational complexity. To validate this expectation, we conducted several numerical experiments detailed before with varying DoFs. To minimize deviations in elapsed time measurements, each experiment was repeated 30 times and the average CPU time was recorded; however, the run-to-run differences were too small to be clearly visualized in the resulting plots.

Fig.~7 presents the results for a two-level L-shaped waveguide discontinuity scenario where \(\mathbbm{u} = {\mathbbm{s}}^{\text{coarse}} + {\mathbbm{s}}^{\text{detail,1}}\). For \(N = 12800\), \(25600\), \(51200\), \(102400\), \(204800\), \(409600\), \(819200\), and \(1638400\) DoFs at the finest level, we measured the CPU time per iteration, \(T\) (in seconds). These measurements include all sparse matrix–vector multiplications and the iterative solution of the linear systems, as described in Algorithm~1 for computing \(\mathbbm{s^{\text{coarse}}}\), \(\mathbbm{s^{\text{detail,1}}}\), and the conventional FEM solution. The slope of the \(\log(T)\) versus \(\log(N)\) plot indicates a computational complexity of approximately \(\mathcal{O}(N^{0.91})\) per iteration. Once \(\mathbbm{s^{\text{coarse}}}\) is computed, finer-level solutions can be incorporated efficiently. In particular, computing \(\mathbbm{s^{\text{detail,1}}}\) costs about \(\mathcal{O}(N^{0.34})\) and can be added to the \(\mathbbm{s^{\text{coarse}}}\) solution, if needed. As anticipated from the theoretical analysis, the \(\mathbbm{s^{\text{coarse}}}\) computation by far dominates the total cost.

Fig.~8(a) presents the logarithmic elapsed time versus DoF graph for a six-level operator-adapted wavelet decomposition-based FEM algorithm applied to the L-shaped waveguide problem. Again, we consider \(N = 12800\), \(25600\), \(51200\), \(102400\), \(204800\), \(409600\), \(819200\), and \(1638400\) DoFs at the finest level, defining the solution as \(\mathbbm{u} = \mathbbm{s^{\text{coarse}}} + \mathbbm{s^{\text{detail,1}}} + \mathbbm{s^{\text{detail,2}}} + \mathbbm{s^{\text{detail,3}}} + \mathbbm{s^{\text{detail,4}}} + \mathbbm{s^{\text{detail,5}}}\). As shown in Fig~8(a), an experimentally measured computational complexity of approximately \(\mathcal{O}(N^{0.92})\) per iteration is obtained. Furthermore, Fig.~8(b) presents the logarithmic elapsed time versus DoF for the same scenario as in Fig.~8(a), now including all precomputation steps, such as the precomputation of \(\tilde{\mathbf{C}}\), \(\tilde{\mathbf{W}}\), and the necessary matrices for creating ILU preconditioners, in addition to the main steps of the algorithm. When precomputations are included, the computational complexity is observed to be approximately \(\mathcal{O}(N^{1.07})\) per iteration. These experimental results validate the nearly linear computational complexity of the entire algorithm. 
As detailed in Section~IV-D, when combined with GMRES and/or LGMRES Krylov subspace iterative solvers and ILU preconditioners, the linear equation solvers
at each level converge within 50–250 iterations for a residual threshold of \(\varepsilon = 10^{-6}\). Across all applications, the number of iterations remains nearly constant and shows minimal dependence on the number of DoFs, owing to the effectiveness of our ILU preconditioner generation sub-algorithm, as described in Section~IV-D. 

The computational complexity trends reported here were consistently observed across all numerical experiments, despite substantial variations in physical dimensions, geometries, and the number of scale levels. For instance, when the L-shaped waveguide discontinuity experiment of Figs.~7 and 8 is repeated over different set of increasingly larger domain sizes (corresponding to $N = 21280$, $34570$, $75740$, $190468$, $311350$, and $2176532$ DoFs), the measured CPU time again exhibits an almost $\mathcal{O}\!\left(N^{0.92}\right)$ computational cost per iteration.

To further substantiate the near-linear scaling claim, Fig.~9 reports CPU time versus DoFs for several representative scenarios.
For example, Fig.~9(b) shows results for the U-shaped waveguide discontinuity over domain sizes yielding
$N = 35450$, $124180$, $347540$, $985740$, and $3681430$ DoFs. As can be seen from the figure, the measured timing data continue to follow the predicted nearly linear computational complexity trend. Additional timing curves obtained for the other numerical experiments, different domain sizes, and scale levels, are also given in Fig.~9.

\begin{figure*}[t]
\centering

\subfloat[]{%
  \includegraphics[width=0.48\linewidth]{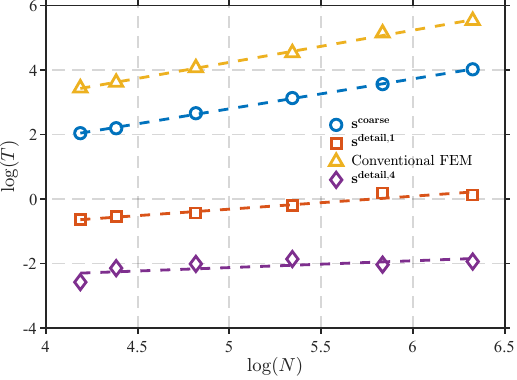}%
  \label{fig:cpu-ushaped-total}
}
\hfill
\subfloat[]{%
  \includegraphics[width=0.48\linewidth]{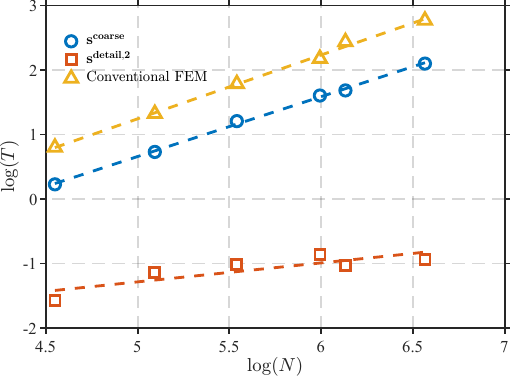}%
  \label{fig:cpu-hollow-total}
}\\[1.0mm]

\subfloat[]{%
  \includegraphics[width=0.48\linewidth]{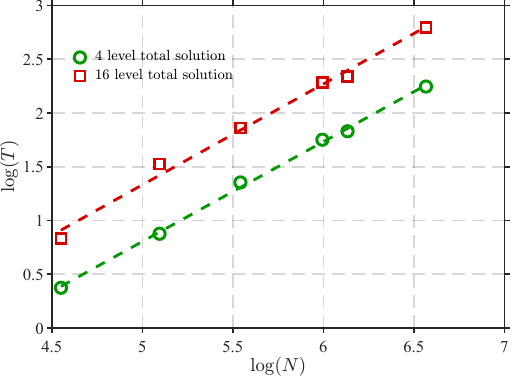}%
  \label{fig:cpu-ushaped-levels}
}
\hfill
\subfloat[]{%
  \includegraphics[width=0.48\linewidth]{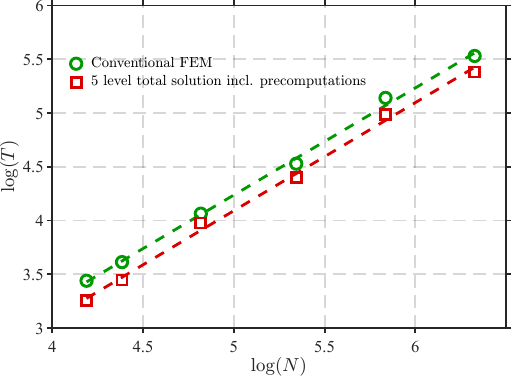}%
  \label{fig:cpu-hollow-levels}
}

\caption{Elapsed CPU time per iteration, $T$, versus degrees of freedom, $N$, in log-log scale for representative numerical experiments solved with the proposed operator-adapted wavelet-decomposition FEM (Algorithm~1). Markers denote measured run-times and dashed lines indicate least-squares fits. (a) Five-level leaky MPSi waveguide: timings for the coarser solution $\mathbbm{s}^{\text{coarse}}$ and representative detail-level solutions $\mathbbm{s}^{\text{detail,1}}$ and $\mathbbm{s}^{\text{detail,4}}$, together with the conventional FEM baseline. (b) Three-level U-shaped waveguide discontinuity: timings for $\mathbbm{s}^{\text{coarse}}$ and $\mathbbm{s}^{\text{detail,2}}$, with the conventional FEM baseline. (c) U-shaped waveguide discontinuity: total-solution run-times for 4-level and 16-level hierarchies. (d) Leaky MPSi waveguide: total-solution runtime for the five-level hierarchy including precomputation steps and calculating all level solutions together with the conventional FEM solution.}

\label{fig:cpu-complexity-2x2}
\end{figure*}

\section{Conclusion and Look Ahead\label{sec:conclusion}}

This study introduced a novel formulation of the operator-adapted wavelet decomposition-based FEM for efficiently solving electromagnetics multiscale EM problems.
The algorithm was constructed hierarchically with near linear computational complexity by using sparse linear-algebra-based techniques across all scale levels. The experimental computational complexity of the main algorithm, including all hierarchical sparse matrix–vector multiplications, was observed to be approximately \(\mathcal{O}(N^{0.91})\) per iteration. With the inclusion of precomputation steps, the computational complexity was observed to be about \(\mathcal{O}(N^{1.07})\) for the examples considered. The precision of the method was consistent with the usual FEM solution at the highest level, achieving a relative \(L^2\) norm error on the order of user-defined thresholds and tolerances.
The proposed algorithm is especially advantageous for analyzing multiscale EM problems because of its ability to eliminate interactions between different scales, allowing independent computation of solutions at each scale level. As a result, finer detail-level solutions could be seamlessly incorporated as needed to improve accuracy. 

Using hybrid and adaptive FEM meshes, the proposed algorithm can be further improved. For example, it can be extended by incorporating arbitrary polygonal mesh elements at coarser levels. In this fashion, regions of the domain with fine geometric details could be captured by finer levels with triangular elements, and smoother regions by larger, coarser-level polygonal elements. Our ongoing work indicates that the computational advantages of the proposed multiscale approach can be further enhanced by employing adaptive mesh hierarchies composed of arbitrary convex polygonal elements together with adaptive coarsening procedures, enabling more efficient treatment of increasingly challenging multiscale problems. In addition, we are extending the proposed method to three dimensions while retaining the core algorithmic structure. In 3-D, the main additional effort lies in the geometric preprocessing required to generate input variables and to construct mesh hierarchies. Once these inputs are available, we anticipate that multiscale solutions can be obtained at nearly linear computational cost. We are also investigating $hp$-refinement strategies employing higher-order basis functions within the same framework.

\FloatBarrier 
\newpage
\appendices
\section{Flowchart of the Algorithm}

Fig.~\ref{fig:oaw-flow} provides a compact overview of the proposed sparse operator-adapted wavelet decomposition-based FEM algorithm (Algorithm~1). The flowchart summarizes (i) the hierarchical construction of the linear operators across scales, (ii) the coarsest-level solve, (iii) the independent detail-level solves up to level $\zeta$ determined by the target accuracy $\beta$, and (iv) the final multilevel solution.

\begin{figure}[h]
\centering
\begin{tikzpicture}[node distance=4.1mm, font=\footnotesize]

  \tikzset{
    flow/.style={
      draw=red!80!black, very thick,
      rounded corners=2mm,
      inner sep=2.6pt,
      align=center,
      fill=white
    },
    flowwide/.style={flow, text width=0.83\linewidth},
    flowformula/.style={flowwide, font=\scriptsize},
    term/.style={flow, ellipse, minimum width=1.65cm, minimum height=6mm},
    decision/.style={
      draw=red!80!black, very thick,
      diamond, aspect=2.15,
      inner sep=1.6pt,
      align=center,
      fill=white
    },
    dblarr/.style={
      double, line width=0.5pt,
      -{Stealth[length=2.2mm]},
      draw=red!80!black
    },
    looparr/.style={dblarr, rounded corners},
    lab/.style={font=\scriptsize, inner sep=1pt},
    grouplab/.style={lab, fill=white, inner sep=1.4pt}
  }

  \node[term] (start) {START};

  \node[flowwide, below=of start] (inputs) {%
    \textbf{Inputs:}\\
    $\{\tilde{\mathbf{C}}^j\}_{j=1}^{q-1}$,\;
    $\{\tilde{\mathbf{W}}^j\}_{j=1}^{q-1}$,\;
    $\tilde{\mathbf{A}}^q=\mathbb{A}^q$,\;
    $\tilde{\bm{\Phi}}^q$,\;
    $\tilde{\mathbf{g}}^q$\\[-0.1em]
    Desired accuracy $\beta$ ($\zeta$ will be determined considering the $\beta$)
  };

  \node[flow, below=of inputs] (initj) {$j\leftarrow q$};
  \node[decision, below=of initj] (condj) {$j\ge 2$?};

  \node[flowformula, below=of condj] (Cdef) {%
    \textbf{Define $\mathbb{C}^{j-1}$ (linear operator):}\\[-0.0em]
    $\begin{aligned}
    \mathbb{C}^{j-1}=
    &\left(\tilde{\mathbf{C}}^{j-1}\tilde{\mathbf{C}}^{j-1,T}\right)^{-1}
      \tilde{\mathbf{C}}^{j-1}\Big[\tilde{\mathbf{I}}-\mathbb{A}^{j}\tilde{\mathbf{W}}^{j-1,T} \cdots \nonumber \\\\
    &\cdots\left(\tilde{\mathbf{W}}^{j-1}\mathbb{A}^{j}\tilde{\mathbf{W}}^{j-1,T}\right)^{-1}
      \tilde{\mathbf{W}}^{j-1}\Big]
    \end{aligned}$
  };

  \node[flowformula, below=of Cdef] (Adef) {%
    \textbf{Define $\mathbb{A}^{j-1}$ (linear operator):}%
    $\;\mathbb{A}^{j-1}=\mathbb{C}^{j-1}\,\mathbb{A}^{j}\,\mathbb{C}^{j-1,T}$
  };

  \node[flow, below=of Adef] (decj) {$j\leftarrow j-1$};

  \node[flowformula, below=of decj] (rhs) {%
    \textbf{Coarsest level RHS:}%
    $\;\mathbbm{g}^{1}=\mathbb{C}^{1}\mathbb{C}^{2}\cdots\mathbb{C}^{q-1}\,\tilde{\mathbf{g}}^{q}$
  };

  \node[flowwide, below=of rhs] (solve1) {%
    \textbf{Coarsest level solve}\\[-0.1em]
    $\mathbb{A}^{1}\mathbbm{v}^{1}=\mathbbm{g}^{1}$\\
    (ILU preconditioned GMRES/LGMRES)
  };

  \node[flowformula, below=of solve1] (scoarse) {%
    \textbf{Coarse-level solution:}
    $\;\mathbbm{s}^{\mathrm{coarse}}=\tilde{\bm{\Phi}}^{q,T}\,\mathbb{C}^{q-1,T}\cdots\allowbreak\mathbb{C}^{1,T}\,\mathbbm{v}^{1}$
  };

  \node[flow, below=of scoarse] (initk) {$j\leftarrow 2$};
  \node[decision, below=of initk] (condk) {$j\le \zeta$?};

  \node[flowformula, below=of condk] (wsolve) {%
    \textbf{Detail coefficients:}\\\
    $\;\tilde{\mathbf{W}}^{j-1}\mathbb{A}^{j}\tilde{\mathbf{W}}^{j-1,T}\,\mathbbm{w}^{j-1}
    =\tilde{\mathbf{W}}^{j-1}\mathbb{C}^{j}\mathbb{C}^{j+1}\cdots\allowbreak\mathbb{C}^{q-1}\,\tilde{\mathbf{g}}^{q}$
  };

  \node[flowformula, below=of wsolve] (sdetail) {%
    \textbf{Detail level solution:}\\\
    $\;\mathbbm{s}^{\mathrm{detail},\,j-1}=\tilde{\bm{\Phi}}^{q,T}\,\mathbb{C}^{q-1,T}\cdots\allowbreak\mathbb{C}^{j,T}\,\tilde{\mathbf{W}}^{j-1,T}\,\mathbbm{w}^{j-1}$
  };

  \node[flow, below=of sdetail] (incj) {$j\leftarrow j+1$};

  \node[flowformula, below=of incj] (assemble) {%
    \textbf{Final solution computation:}%
    $\;\mathbbm{u}^{\zeta}=\mathbbm{s}^{\mathrm{coarse}}
      +\sum_{j=1}^{\zeta-1}\mathbbm{s}^{\mathrm{detail},\,j}$
  };

  \node[term, below=of assemble] (end) {END};

  \begin{pgfonlayer}{background}
    \node[
      draw=red!80!black, dashed, rounded corners=2mm, inner sep=5pt,
      fit=(initj)(condj)(Cdef)(Adef)(decj),
      name=opbox
    ] {};
    \node[grouplab, anchor=north west]
      at ([xshift=2pt,yshift=-2pt]opbox.north west)
      {Linear operator construction};

    \node[
      draw=red!80!black, dashed, rounded corners=2mm, inner sep=5pt,
      fit=(initk)(condk)(wsolve)(sdetail)(incj),
      name=detbox
    ] {};
    \node[grouplab, anchor=north west]
      at ([xshift=2pt,yshift=-2pt]detbox.north west)
      {Detail level solves};
  \end{pgfonlayer}

  \draw[dblarr] (start) -- (inputs);
  \draw[dblarr] (inputs) -- (initj);
  \draw[dblarr] (initj) -- (condj);

  \draw[dblarr] (condj) -- node[lab,right] {Yes} (Cdef);
  \draw[dblarr] (Cdef) -- (Adef);
  \draw[dblarr] (Adef) -- (decj);

  \draw[looparr]
    (decj.west) --
    ([xshift=-3mm]Cdef.west |- decj.west) |-
    (condj.west);

  \draw[looparr]
    (condj.east) --
    ([xshift=3mm]Cdef.east |- condj.east) |-
    node[lab,right,pos=0.25] {No}
    (rhs.east);

  \draw[dblarr] (rhs) -- (solve1);
  \draw[dblarr] (solve1) -- (scoarse);
  \draw[dblarr] (scoarse) -- (initk);
  \draw[dblarr] (initk) -- (condk);

  \draw[dblarr] (condk) -- node[lab,right] {Yes} (wsolve);
  \draw[dblarr] (wsolve) -- (sdetail);
  \draw[dblarr] (sdetail) -- (incj);

  \draw[looparr]
    (incj.west) --
    ([xshift=-3mm]wsolve.west |- incj.west) |-
    (condk.west);

  \draw[looparr]
    (condk.east) --
    ([xshift=3mm]wsolve.east |- condk.east) |-
    node[lab,right,pos=0.25] {No}
    (assemble.east);

  \draw[dblarr] (assemble) -- (end);

\end{tikzpicture}

\caption{Flowchart summarizing the proposed sparse operator-adapted wavelet decomposition-based FEM algorithm (Algorithm~1).}

\label{fig:oaw-flow}
\end{figure}

\FloatBarrier

\ifCLASSOPTIONcaptionsoff
  \newpage
\fi



%

%
\bibliographystyle{ieeetr}
\bibliography{main}



\end{document}